\RequirePackage{fix-cm}

 \documentclass[showkeys,preprint]{revtex4}          % twocolumn%
\usepackage{graphicx}
\usepackage{amssymb}
\usepackage{amsbsy}
\usepackage{subfigure}
\usepackage{color}
\usepackage{enumitem}

\newtheorem{algo}{Algorithm}

\begin{document}

\title[]{Measure of combined effects of morphological parameters %and the random distribution 
of inclusions 
within composite materials via stochastic homogenization to determine 
effective mechanical properties
}

\author{Vladimir Salnikov} 
\author{Sophie Lemaitre}
\author{Daniel Cho\"i}
\author{Philippe Karamian-Surville}

\affiliation{Nicolas Oresme Mathematics Laboratory \\ University of Caen Lower Normandy\\ 
  CS 14032, Bd. Mar\'echal Juin,  BP 5186\\
  14032, Caen Cedex,  France \\
  \email{vladimir.salnikov@unicaen.fr, daniel.choi@unicaen.fr, philippe.karamian@unicaen.fr}
}


%\date{19 September 2014}

\begin{abstract}
In our previous papers we have described efficient and reliable methods of 
generation of representative volume elements (RVE) perfectly suitable for 
analysis of composite materials via stochastic homogenization. 

In this paper we profit from these methods to analyze the influence of the morphology on 
the effective mechanical properties of the samples. 
More precisely, we study the dependence of main mechanical characteristics of 
a composite medium on various parameters of the mixture of inclusions composed of spheres 
and cylinders. On top of that we introduce various imperfections to 
inclusions and observe the evolution of effective properties related to that. 

The main computational approach used throughout the work is the FFT-based homogenization
technique, validated however by comparison with the direct finite elements method. 
We give details on the features of the method and the validation campaign as well.


%  \subclass{MSC code1 \and MSC code2 \and more}
\end{abstract}

\maketitle

\section{Introduction / motivation}
\label{sec:intro}
In this paper we study the influence of morphological parameters of composite 
materials on their effective mechanical properties. The usage of composite materials 
for industrial applications motivated a huge amount of publications on the subject 
in recent years: they concern both experimental and modelling results. The reason for us 
to address the question is twofold as well: on the one hand we explore 
the existing modelling techniques, on the other hand we have in mind very concrete 
applications related to the project in the industry of aeronautics.

The need in modelling for the analysis of composite materials as in most of the applied domains 
comes from the fact that experimental work is usually expensive and difficult to carry out. 
It is thus important to develop modelling approaches that are efficient, reliable and 
sufficiently flexible so that the outcome can be validated by an experiment. 
The strategy that we adopt here is related to the notions of stochastic 
homogenization. The key idea is to consider a sample of a composite material 
that is sufficiently large to capture its behavior and compute the macroscopic parameters: 
for mechanical properties it can be for example the Young modulus, the Poisson ratio 
or eventually the whole stiffness tensor. To take into account possible imperfections 
or random factors one can average the result for a series of tests representing the 
same macro characteristics. The usual technology for this is to generate a series of samples
(representative volume elements) randomly, 
controlling though their parameters, perform the computation for each of them and average the
result.

It is now generally accepted that the main characteristic affecting the effective properties 
of a composite material is its morphology, i.e. the combination of geometric characteristics 
of the inclusions and their distribution in the supporting matrix. To analyze the phenomenon 
one needs thus a tool to generate RVEs capturing various morphological parameters. 
We have developed and implemented such a tool: in \cite{VDP} we described the algorithms to 
produce the RVEs containing spheres (that represent globular inclusions) and cylinders (that are 
responsible for fiber-type reinforcements). We are able to reach the volume fraction of inclusions 
up to relatively high values of 50\%--60\%, and in addition we can control the geometric configuration of 
a sample as a whole, namely manage the intersections of inclusions and eventually their distribution. 
Moreover in \cite{SVDP} 
we have extended the method to introduce irregularities to the shape of inclusions. 
In this paper we describe the results of computations carried out with the generated samples 
via an FFT-based homogenization technique (\cite{moul-suq, monchiet}). 
For the presentation here we have chosen the results that can be useful for applications 
and/or those where the trends are not intuitively obvious, in particular we explore the 
influence of redistribution of the volume fraction between globular and fiber-type reinforcements, 
as well as the effects of imperfections. 

The paper is organized as follows.
In the next section we briefly recall the RVE generation methods proposed in \cite{VDP}
and extended in our further works. % \cite{SVDP}. 
In its second part %(\ref{sec:FFT}) 
we give details about the main computational method which is used throughout the work: 
the FFT-based homogenization technique coupled with stochastic methods of RVE generation. 
We describe its convenience and limitations as well as present the results of the validation 
campaign, i.e. compare it with the direct finite elements method. 
The section \ref{sec:mech} is the description of the results of analysis (via the above 
mentioned methods) of effective properties of composites depending on the morphological parameters 
of the inclusions. We conclude by describing eventual industrial applications, some work in progress
and expected results of it.

\section{Sample generation and computational techniques} 
As we have outlined in the introduction, this section is devoted to a brief description of the 
methods that we have used to perform computations and the reasons to choose these concrete methods.

\subsection{RVE generation} \label{sec:RVE}
The generation of samples for computation is an important step in the process of modelling
of the behavior of composite materials. Since the morphology of composites may be 
quite complex this can be a very challenging task. On the one hand it is important to 
be able to approximate rather involved geometries, on the other hand the method should be 
fast and reliable; in the ideal case the stage of generation should be much shorter than 
the computation itself. 

There has been a number of works where the inclusions were 
represented by simple geometric objects like spheres or ellipsoids 
(see for example, \cite{segu, levesque_sp, zhao, man}). If one considers more 
complicated geometry, the problem of managing the intersection of inclusions 
arises immediately. Among the established approaches of dealing with it, one can 
mention two important families: random sequential adsorption (RSA) type algorithms and 
molecular dynamics (MD) based methods.
The RSA (\cite{rsa1}) is based on sequential addition of inclusions verifying for each of them 
the intersection; the main idea of the MD (\cite{LSA, williams, levesque_ell}) is to make the inclusions move, 
until they reach the desired configuration. Let us mention that the first method needs an efficient 
algorithm of verification of intersection between the geometric shapes, and the second one 
an algorithm of predicting the time to the intersection of moving objects, which exists 
for a very limited class of shapes and often amounts to a difficult minimization problem. 
In \cite{VDP}, we have described the classical RSA and a time-driven version of MD applied to the 
mixture of inclusions of spherical and cylindrical shapes.
The key ingredient for both of the approaches was the 
explicit formulation of algebraic conditions of intersection of a cylinder with a sphere and of two 
cylinders. 
To be more specific, we recapitulate the ideas of these algorithms (Algorithms \ref{alg:mc_gen}, \ref{alg:md_gen}).

\begin{figure}[ht]  
\centering
 \includegraphics[width=0.51\linewidth]{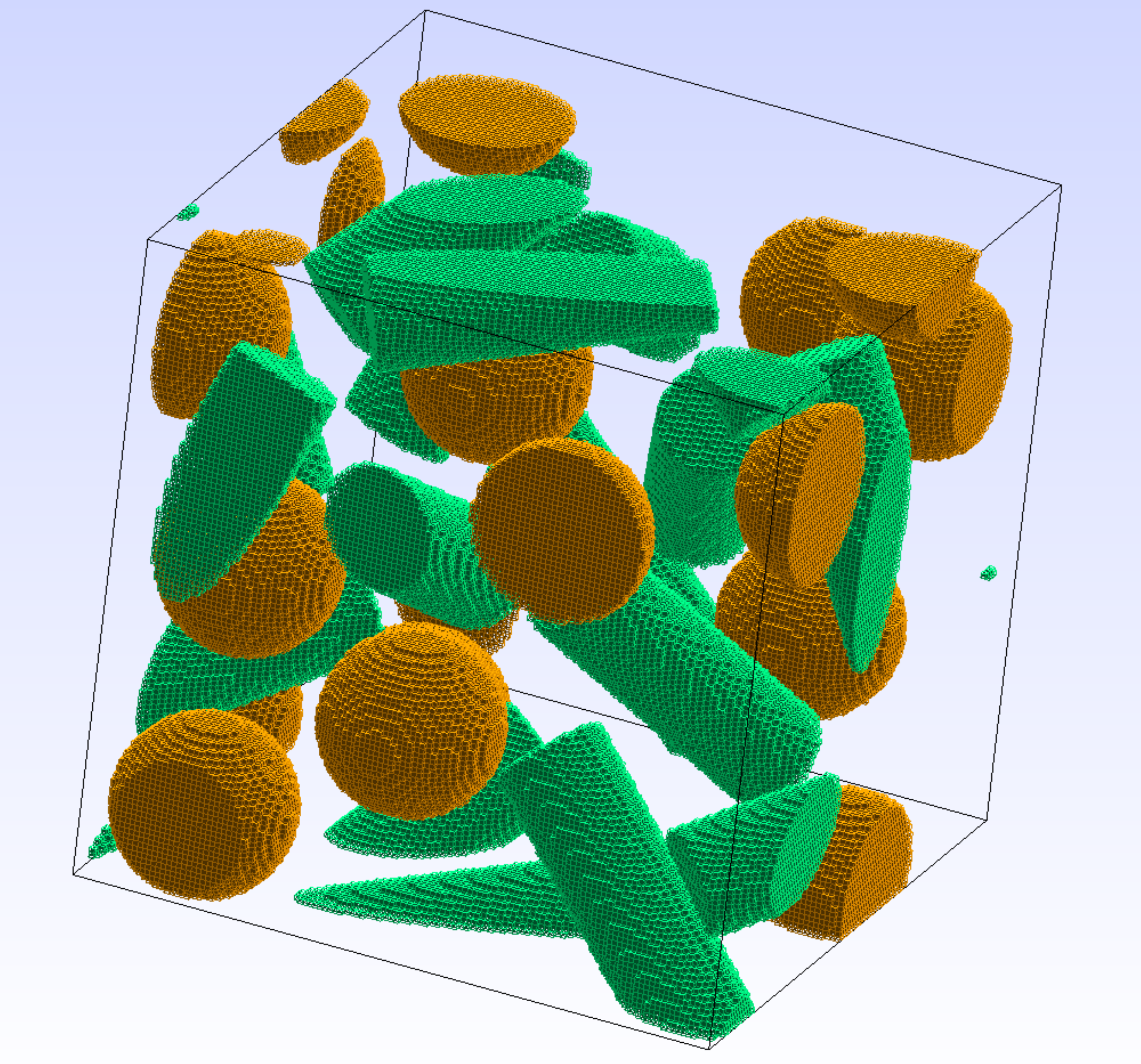} 
\caption{\label{fig:rve3D} 3D view of an RVE: spherical and cylindrical inclusions, periodic boundary 
conditions.} 
 \end{figure}

We have observed that the RSA approach is extremely efficient for relatively small 
volume fractions of inclusions (up to $30 \%$), where it permits to generate a sample 
in fractions of a second. The MD-based method is powerful for higher volume fractions (of order $50 -60 \%$):
it generates a configuration in about a second while the RSA can get stuck. An example of a sample with a mixture 
of non-intersecting spherical and cylindrical inclusions is presented on the figure \ref{fig:rve3D}.

\begin{samepage}
\begin{algo} \label{alg:mc_gen} \underline{RSA generation procedure}. \newline
Input: volume fractions, number of inclusions $n_{cyl}, n_{sp}$, aspect ratio.
\newline \begin{tabular}{c|l}   & \parbox{0.95\linewidth}{
\vspace{-0.5em}\begin{enumerate}[itemsep=-0.2em]
 \item  Compute the parameters of cylinders and spheres 
 \item  NumOfGenSpheres = 0,  NumOfGenCylinders = 0
 \item  while $(NumOfGenCylinders < n_{cyl})$
      \newline \begin{tabular}{c|l}   & \parbox{0.9\linewidth}{
      \begin{enumerate}[itemsep=-0.2em]
       \item Generate a new cylinder \label{gencyl} 
       \item Using the algorithm (\cite{VDP}, Alg. 4) check if it overlaps with any cylinder generated before
       \item if yes redo  \ref{gencyl}
       \item if no increase NumOfGenCylinders
       \end{enumerate}
        }\end{tabular}
 \item  while $(NumOfGenSpheres < n_{sp})$
      \newline \begin{tabular}{c|l}   & \parbox{0.9\linewidth}{
           \begin{enumerate}[itemsep=-0.2em]
       \item Generate a new sphere \label{gensp} 
       \item Check if it overlaps with any sphere generated before
       \item if yes redo \ref{gensp}
       \item Using the algorithm (\cite{VDP}, Alg. 1) check if it overlaps with any cylinder generated before
       \item if yes redo \ref{gensp}
       \item if no increase NumOfGenSpheres
       \end{enumerate}
        }\end{tabular}
  
\end{enumerate}
}\end{tabular}
 
\end{algo}
\end{samepage}

\begin{samepage}
\begin{algo} \label{alg:md_gen} \underline{MD-based generation procedure}. \newline
Input: volume fractions, number of inclusions $n_{cyl}, n_{sp}$, aspect ratio.
\newline \begin{tabular}{c|l}   & \parbox{0.95\linewidth}{
\vspace{-0.5em}\begin{enumerate}[itemsep=-0.25em]
 \item  Compute the parameters of cylinders and spheres, fix the criterion $\varepsilon$ to stop the 
 simulation
 \item  Generate $n_{sp}$ spheres and $n_{cyl}$ cylinders, disregarding overlapping
 \item  while (OverlappingEnergy $> \varepsilon$)
      \newline \begin{tabular}{c|l}   & \parbox{0.9\linewidth}{
      \begin{enumerate}[itemsep=-0.2em]
       \item Check overlapping (\cite{VDP}, Alg. 1, 4)
       \item For each couple of overlapping inclusions compute the interaction force (\cite{VDP}, table 1)
       \item Perform the integration step for the evolution equations (\cite{VDP}, Eq. 6)
       \item Update the value of OverlappingEnergy
       \end{enumerate}
        }\end{tabular}
\end{enumerate}
}\end{tabular}
 
\end{algo}
\end{samepage}
\newpage

\begin{figure}[ht]  
\centering
\subfigure[\, Waved surface of an inclusion]{    
 \includegraphics[height=0.37\linewidth]{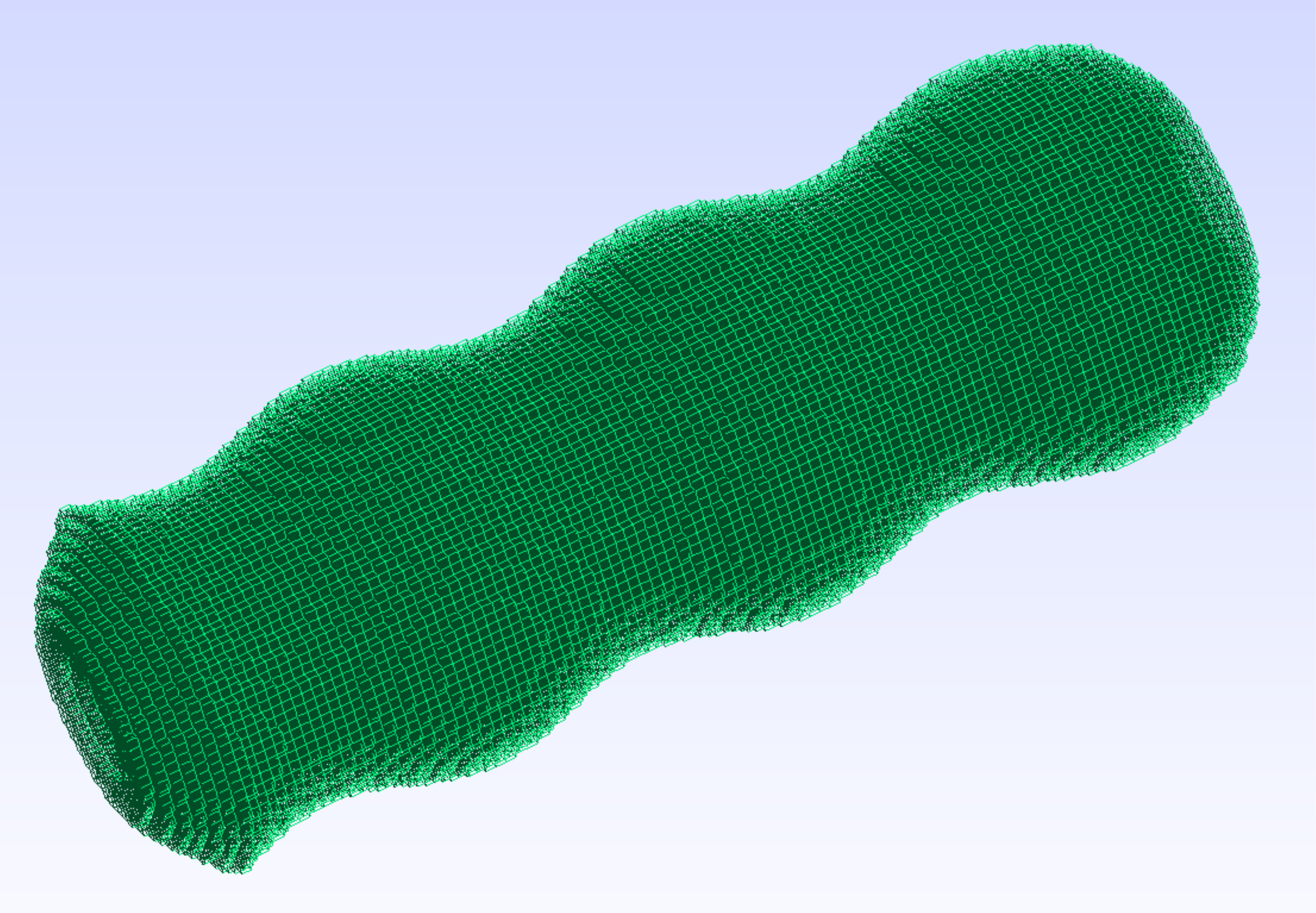} 
}
\subfigure[\, Part of an inclusion taken out]{ 
   \includegraphics[height=0.37\linewidth]{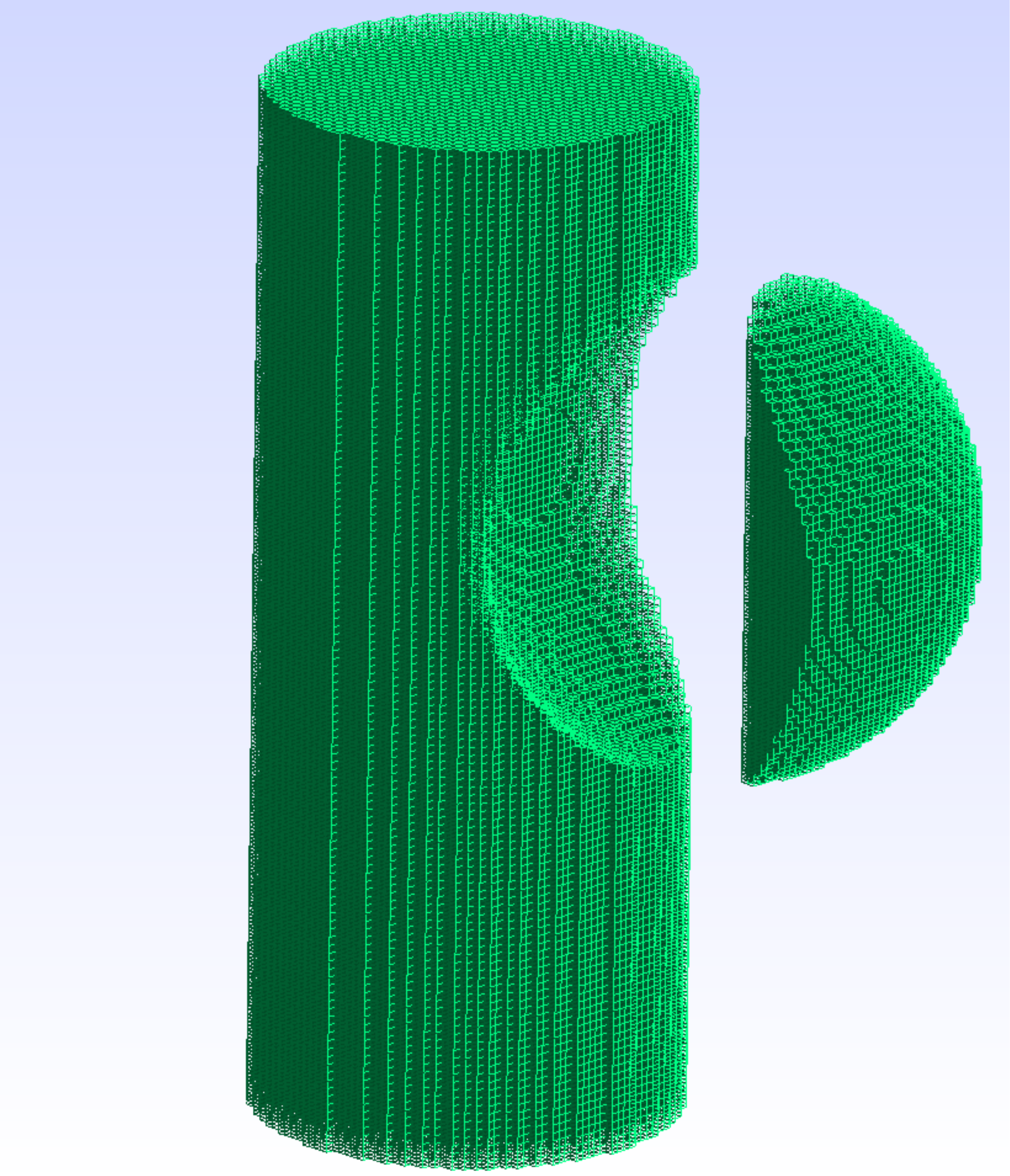}
 }
\caption{\label{fig:imperfections} Various imperfections of inclusions.
} 
 \end{figure}

The outcome of these algorithms is a list of inclusions in the ``vector'' form, i.e. 
a list of coordinates of centers, radii, and eventually axes of symmetry of inclusions. 
This is perfectly suitable for various computational techniques: 
FFT-based homogenization procedures applied to the pixelized samples, as well as finite element 
computations on the mesh constructed from this pixelization. In addition (\cite{SVDP}) we are able to 
introduce various imperfections of the inclusions without spoiling the efficiency of the generation 
algorithm: the figure \ref{fig:imperfections} shows two such examples, where the surface of an inclusion 
is waved or a part of an inclusion is taken out to produce irregular shapes.

\subsection{FFT-based homogenization scheme} \label{sec:FFT}
Having an efficient scheme of generation of samples we can now proceed to the computation of 
effective mechanical properties. As we have mentioned in the introduction for evaluating these 
properties we have adopted the philosophy of stochastic homogenization.
Schematically one can view the process as follows: \\[-1.9em]
\begin{enumerate}[leftmargin= 1.3em,  itemsep=-0.5em]
 \item Fix the macroscopic parameters of the 
material (volume fraction and type of inclusions).
 \item Generate a series of samples (RVEs) of a composite material with these parameters \\[-0.5em]
 (\emph{stochastic} part).
 \item Perform accurate computation of effective properties on these RVEs  \\[-0.5em]
 (deterministic \emph{homogenization} part).
 \item Average the computed (macroscopic) characteristics of the samples.
\end{enumerate}

Let us now describe the mechanical model behind this computation as well as the main 
computational method -- the homogenization procedure. 

Consider a representative volume element $V$, and denote $\mathbf{u}(\mathbf{x})$  the displacement 
field defined at any point $\mathbf{x} \in V$. The system in mechanical equilibrium 
is described by the law $\sigma (\mathbf{x}) = \frac{\partial w(\mathbf{x})}{\partial\varepsilon(\mathbf{x})}$,
where $\varepsilon(\mathbf{x}) = \varepsilon(\mathbf{u}(\mathbf{x})) = 
\frac{1}{2}(\nabla\mathbf{u}(\mathbf{x}) + \nabla\mathbf{u}(\mathbf{x})^T)$ -- the strain tensor in the model of small deformations, 
$w(\mathbf{x})$ -- the stored mechanical energy, 
and $\sigma(\mathbf{x})$ -- the stress tensor, subject to the condition $\mathrm{div}\, \sigma(\mathbf{x}) = 0$.
In the linear case this law simplifies to 
$\frac{\partial w(\mathbf{x})}{\partial\varepsilon(\mathbf{x})} = c(\mathbf{x})\!:\!\varepsilon(\mathbf{x})$
with the stiffness tensor $c(\mathbf{x})$.
Notice that for a composite material the stiffness tensor does depend on the point $\mathbf{x}$: 
the dependence is governed by microscopic geometry of the sample, namely which phase (matrix or inclusion)
the point $\mathbf{x}$ belongs to.
We suppose that the averaged strain $<\!\varepsilon\!> = E$ is prescribed, and decompose $\varepsilon(\mathbf{x})$ 
in two parts: $\varepsilon(\mathbf{u}(\mathbf{x})) = E + \varepsilon(\mathbf{\tilde u}(\mathbf{x}))$, which is 
equivalent to representing  $\mathbf{u}(\mathbf{x}) = E.\mathbf{x} + \mathbf{\tilde u}(\mathbf{x})$, 
for $\mathbf{\tilde u}(\mathbf{x})$ being periodic on the boundary of $V$.  
Thus, the problem we are actually solving reads
\begin{eqnarray} \label{pb1}
  \sigma (\mathbf{x}) = c(\mathbf{x}) : (E + \varepsilon(\mathbf{\tilde u}(\mathbf{x})), 
  \quad \mathrm{div}  \, \sigma(\mathbf{x}) = 0, \\
  \mathbf{\tilde u}(\mathbf{x}) \,\, \mathrm{ periodic}, \, 
  {\sigma}(\mathbf{x}).\mathbf{n} \,\,\mathrm{ antiperiodic}.\nonumber 
\end{eqnarray}
The solution of (\ref{pb1}) is the tensor field $\sigma(\mathbf{x})$, we are interested in its average 
in order to obtain the homogenized stiffness tensor $c_{hom}$ from the equation 
\begin{equation}\label{average}
<\!\sigma(\mathbf{x})\!> = c_{hom} : <\!\varepsilon(\mathbf{x})\!>.
\end{equation}
To recover all the components 
of $c_{hom}$ in 3-dimensional space one needs to perform the computation of $<\!\sigma(\mathbf{x})\!>$ for six 
independent deformations $E$, which morally correspond to usual stretch and shear tests. 

There is a couple of natural approaches to solving the problem (\ref{pb1}): one can construct a 
mesh of an RVE $V$ and employ the finite elements method, or discretize (basically pixelize) the RVE to use 
the FFT-based homogenization scheme (\cite{moul-suq, monchiet}). 
The major difficulty arising when applying the former method is that for rather involved 
geometry one needs to construct a fine mesh which is a non-trivial task in its own, 
and moreover to proceed with finite elements one requires considerable memory resources. 
The idea of the latter 
is that in the Fourier space the equations of (\ref{pb1}) acquire a rather nice form
for which in the case of a homogeneous isotropic material one can construct a Green operator 
and basically produce an exact solution. For a composite material containing possibly 
several phases one introduces an artificial reference medium for which the Green operator
is defined, the computation then is an iterative procedure to approximate the corrections
of the microscopic behavior of the material in comparison to this reference medium.
For the sake of completeness let us present this method here in details, following essentially 
the works \cite{MMS, eyre}.

Introduce the reference medium stiffness tensor $c^0$ and the correction to it  \newline
$\delta c(\mathbf{x}) = c(\mathbf{x}) - c^0$. The equations (\ref{pb1}) can equivalently be rewritten as 
\begin{eqnarray} \label{pb2}
  \sigma (\mathbf{x}) = c^0 : \varepsilon(\mathbf{\tilde u}(\mathbf{x})) + \tau(\mathbf{x}), 
  \quad \mathrm{div}  \, \sigma(\mathbf{x}) = 0, \\
  \tau(\mathbf{x}) =  \delta c (\mathbf{x}) : (\varepsilon(\mathbf{\tilde u}(\mathbf{x})) + E) + c^0 : E, \nonumber \\ 
    \mathbf{\tilde u}(\mathbf{x}) \,\, \mathrm{ periodic}, \, 
  {\sigma}(\mathbf{x}).\mathbf{n} \,\,\mathrm{ antiperiodic}. \nonumber  
\end{eqnarray}
The tensor $\tau$ is called polarization.
The periodicity assumptions permit 
to rewrite the first line of (\ref{pb2}) in the Fourier space. Using the linearity of the Fourier transform and its property
with respect to derivation, one obtains
\begin{equation} \label{pb3}
  \hat \sigma_{mj} (\mathbf{\xi}) = i c^0_{mjkl} \xi_l \hat{\tilde u}_k (\mathbf{\xi}) + \hat\tau_{mj} (\mathbf{\xi}), 
  \quad i \hat \sigma_{mj} \xi_j  = 0,    \nonumber
\end{equation}
where $\hat{\bullet}$ denotes the Fourier image of $\bullet$ and $\xi_j$'s are the coordinates in the Fourier space. 
The key observation is that in the Fourier space there is a relation between the polarization
and the deformation tensors, namely  \newline
$\hat{\tilde{\varepsilon}} (\mathbf{\xi}) = - \hat \Gamma^0(\mathbf{\xi}) : \hat \tau (\mathbf{\xi})$.
$\Gamma^0$ is the  \emph{Green operator}, which for an isotropic reference medium with 
the Lam\'e coefficients $\lambda_0, \mu_0$ can be computed explicitly:
$$
  \hat \Gamma^0_{klmj}(\mathbf{\xi}) = \frac{\displaystyle \delta_{km}\xi_l \xi_j + \delta_{lm}\xi_k \xi_j + \delta_{kj}\xi_l \xi_m + \delta_{lj}\xi_k \xi_m}{\displaystyle 4 \mu_0 |\mathbf{\xi}|^2} 
    -  \frac{\displaystyle \lambda_0 + \mu_0}{\displaystyle \mu_0(\lambda_0 + 2\mu_0)}\frac{\displaystyle  \xi_j \xi_k \xi_l \xi_m}{\displaystyle |\mathbf{\xi}|^4}
$$
Going back to the original variables, the initial problem (\ref{pb2})
reduces to the periodic \emph{Lippmann--Schwinger integral equation} 
$$
  \varepsilon(\mathbf{u}(\mathbf{x})) = - \Gamma^0(\delta c : \varepsilon(\mathbf{u}(\mathbf{x}))) + E,
$$
This equation can be solved iteratively using the following algorithm.
\begin{samepage}
\begin{algo} \label{alg:acc} \underline{FFT-based numerical scheme}.  \newline
Initialize $\varepsilon^0(\mathbf x) \equiv E$, fix the convergence criterion $acc$. \newline
while (not converged)
\newline \begin{tabular}{c|l}   & \parbox{0.95\linewidth}{
\vspace{-0.5em}\begin{enumerate}
 \item Convergence test:  \newline
    if $(\epsilon_{comp}< acc)$
    compute $\sigma(\mathbf{x}) = c(\mathbf{x}):\varepsilon^n(\mathbf{x})$, $\hat \sigma(\mathbf{\xi}) = FFT(\sigma)$, 
    \newline $\epsilon_{eq} = \sqrt{<\!\| \mathbf{\xi}  \hat \sigma(\mathbf{\xi})^n \|^2\!>} /\| \hat \sigma(0)\|$ 
    \newline if $(\epsilon_{eq} < acc) \rightarrow converged, stop$
 %\item  $\sigma^n(\mathbf x) = c(\mathbf x) : \varepsilon(\mathbf x)$
 \item  $\tau^n(\mathbf x) = (c(\mathbf x) + c^0) : \varepsilon^n(\mathbf x) $
 \item $\hat \tau^n(\mathbf{\xi}) = FFT(\tau^n)$
 \item $\hat \varepsilon^n_{comp}(\mathbf{\xi}) = \hat \Gamma^0(\mathbf{\xi}) : \hat \tau^n(\mathbf{\xi}) $, 
 $\mathbf{\xi} \neq 0$; $\quad$  $\hat \varepsilon^n_{comp}(0) = E$
 \item $\varepsilon^n_{comp}(\mathbf{x}) = FFT^{-1}(\hat \varepsilon^n_{comp})$
 \item $\epsilon_{comp} = \sqrt{<\!\|\varepsilon^n - \varepsilon^n_{comp} \|^2\!>} /\|E\|$
 \item $\varepsilon^{n+1}(\mathbf{x}) = \varepsilon^n(\mathbf{x}) - 2(c(\mathbf{x}) - c^0)^{-1}:c^0:(\varepsilon^n_{comp}(\mathbf{x}) - \varepsilon^n(\mathbf{x}))$
\end{enumerate}
}\end{tabular}
 
\end{algo}
\end{samepage}
When the above algorithm converges we can compute $<\!\sigma(\mathbf{x})\!>$ to be inserted into the equation 
(\ref{average}). As we have mentioned above this computation has to be repeated for a complete set of 
independent global deformation fields.

We have chosen this approach for its computational efficiency both in time and memory consumption, 
and also for its convenience in the applied problems, that we will discuss afterwards. 
There are however some details worth being commented on here. 

First, there are several versions of such FFT-based schemes: there is a possibility to use 
the initial variables $\sigma$ and $\varepsilon$ or formulate a dual problem the solution 
of which will produce the compliance tensor. One can produce a sort of mixture of the two, 
using polarization as the primary variable. The efficiency of these methods depends on the 
contrast between the characteristics of different phases. We have chosen to implement 
the direct accelerated scheme presented above, since there one has a theoretical 
result on the optimal values of parameters of the reference medium. According to the convergence 
theorem formulated in \cite{eyre}, for a two-phase composite with the Lam\'e
coefficients of the constituent phases $(\lambda_1, \mu_1)$ and $(\lambda_2, \mu_2)$
respectively, one should choose $\lambda_0 = -\sqrt{\lambda_1 \lambda_2}$,
$\mu_0 = -\sqrt{\mu_1 \mu_2}$ to compute $c^0$.
The optimality of this accelerated scheme is also coherent with recent 
results of \cite{2014}.

Second, we have carried out a validation campaign by comparing the results of computations 
using the FFT-based process, several types of finite elements approaches, and analytical results
where possible. 
We have constructed some test samples with simple geometries: a square bar with 
different orientations, a plane cutting the sample in two parts, etc. In addition to the FFT-based 
scheme for these samples we have computed the stiffness tensor using finite elements with the adapted mesh 
and with the mesh constructed from the pixelized sample. 
In both cases we used about $2^{18}$ hexahedron elements : parallelepipeds for the adapted mesh 
constructed with Cast3M, and cubes corresponding to voxels for the mesh built from 
pixelized sample.
The computation with adapted meshes gives results
which are in good agreement with theoretical estimations, while with the mesh obtained from the pixelization
there is a tendency of overestimating the parameters (when the inclusions are more rigid than the matrix).
Although for the values of contrast between the matrix and the inclusions that interest us 
the difference can reach $1-2\%$, which is not of great importance for observing the trends. 
What is more interesting is that the FFT-based scheme produces the results where the parameters are 
often underestimated (up to $5\%$). The situation is certainly reversed when 
the inclusions are less rigid than the matrix. For computing mechanical properties of
composite materials a simple way out is to increase the resolution of pixelization. 
For the tests that we describe in this paper the pixelization at around $200\times200\times200$
already produces reasonable results. For each given sample like the one presented at figure 
\ref{fig:rve3D} it is sufficient to make several tests at different resolutions for the same 
vector data and observe when the result stabilizes.
Let us however note that if one is interested in thermal properties, where typically 
there are more than two phases with rather fine geometry,
the problem can be much more complicated -- 
we will suggest some more advanced techniques of adapting a sample to such computations elsewhere. 

Third, as mentioned the stochastic part of computations is due to averaging the results 
for several samples with the same macroscopic characteristics. 
This is the usual approach inspired by the Monte Carlo method. Since the pioneer work \cite{monte-carlo}
of Metropolis and Ulam, there has been a great number of applications of this method for various problems in 
pure mathematics (\cite{VS}), physics (\cite{kats}), engineering (\cite{mc-eng, levesque_ell, Leclerc2012, Leclerc2013b, Leclerc2013c}), science in 
general\footnote{This list is in no case exhaustive since the range of application of the Monte Carlo approach 
is, without exaggeration, enormous.}. 
In the concrete context of homogenization and estimation of effective properties of composite materials
a natural problem arises: having computed the average for a given sampling we need to decide how far 
it is from the real average. The usual way to do it is to use Student's distribution 
to compute the confidence intervals from the sampling mean value and the standard deviation
(see for example \cite{levesque_sp}) for a precise algorithm. In our tests it was sufficient 
to make $10 - 20$ runs to obtain acceptably small values of deviation, and $20$ was needed 
not very often -- mostly for high values of volume fraction of inclusions and significant contrast 
between the two materials. We do not depict the confidence intervals on the figures that follow 
not to overload the plots, we however verify for each of them that the trends that we exhibit  
are not due to statistical errors.

\section{Mechanical properties of composites} \label{sec:mech}
In this section we present the tendencies of the behaviour of 
composite materials for various morphological parameters, 
that we have obtained by making a series of tests using the algorithms 
described above. We start with simple tests, the aim of which is more to 
validate the methods in the sense that they produce expected 
results for simple tendencies. We continue with more involved 
analysis of influence of combinations of morphological parameters.

With the algorithm of generation of RVEs we are able to control a number of parameters.
As we have already mentioned we are working with spheres and cylinders that are 
supposed to represent respectively globular inclusions and microfiber reinforcements. 
For both types of inclusions we are able to assign the volume fraction in the 
generated sample: $f_{sp}$ and $f_{cyl}$ respectively. We can choose the number of inclusions for each type
($n_{sp}$, $n_{cyl}$), and for 
cylinders we have an extra parameter of aspect ratio (the ratio between the 
length of a cylinder and its diameter). This already gives a lot of parameters, 
on top of that we will introduce the imperfections to inclusions -- we will comment 
on them at the end of this section. 
The result certainly depends on the  mechanical parameters of the 
matrix and the inclusions -- we describe this concisely by fixing the value of contrast between 
the two media. The output is thus the normalized (with respect to the matrix) value of 
homogenized parameters. We have carried out several series\footnote{This is a rather time consuming 
numerical experiment, that took about 500Khours of CPU time at the Antares cluster of the CRIHAN computational center.} 
of computations varying these parameters, in this paper we present a selection of results that are 
qualitatively not obvious from the first sight or quantitatively important for applications.

\subsection{Basic tests} 
\begin{figure}[ht]  
\centering
 \includegraphics[width=0.8\linewidth]{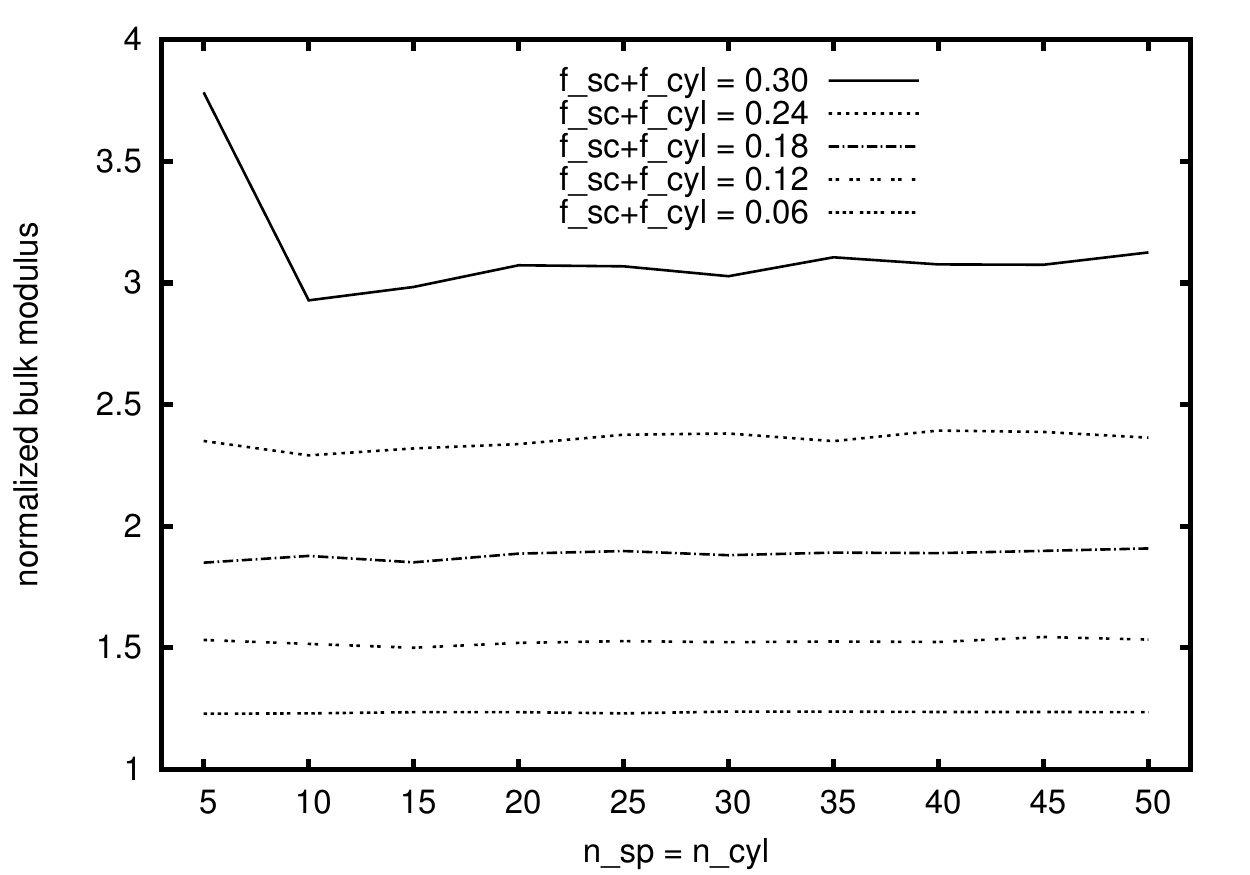} 

\caption{\label{fig:rve} The dependence of mechanical parameters of a composite
on the number of inclusions for the volume fraction being fixed. 
Contrast 2048. 
} 
 \end{figure}

Before starting the real computations we need to fix one more detail, namely the typical size of the representative volume 
element. According to \cite{kanit} the size is acceptable if increasing it 
does not modify the result of computations. In our case fixing the relative size 
of an RVE is equivalent to determine the minimal acceptable number of inclusions. 
A small number of inclusions clearly corresponds to a small piece of material 
studied, while the large number means that each sample includes sufficient 
microscopic variety and is close to being homogeneous in the context of effective properties. 
We have analyzed the dependence of effective properties on the number of inclusions 
with all the other parameters being fixed. The typical trend is shown on the figure \ref{fig:rve}, 
which clearly shows that from the total number of $20$ the properties stabilize, the effect is more pronounced 
for higher volume fractions.
In all the computations that follow, we will thus consider the number of inclusions of this order.

Let us now turn to analysis of the influence of morphological properties. 
Here and in what follows 
we will depict the trends for the effective bulk and shear moduli 
or for just one of them, since they are usually rather similar qualitatively.
In principal we are computing the whole homogenized stiffness tensor, that turns out to 
be very close to isotropic, it is thus possible to extract any combination 
of mechanical characteristics out of it (Young modulus, Poisson ratio, Lam\'e coefficients).
The choice of the bulk and the shear moduli is motivated by further comparison with 
experimental measurements and datasheets.

\begin{figure}[ht]  
\centering
\subfigure[\, Bulk modulus]{    
 \includegraphics[width=0.45\linewidth]{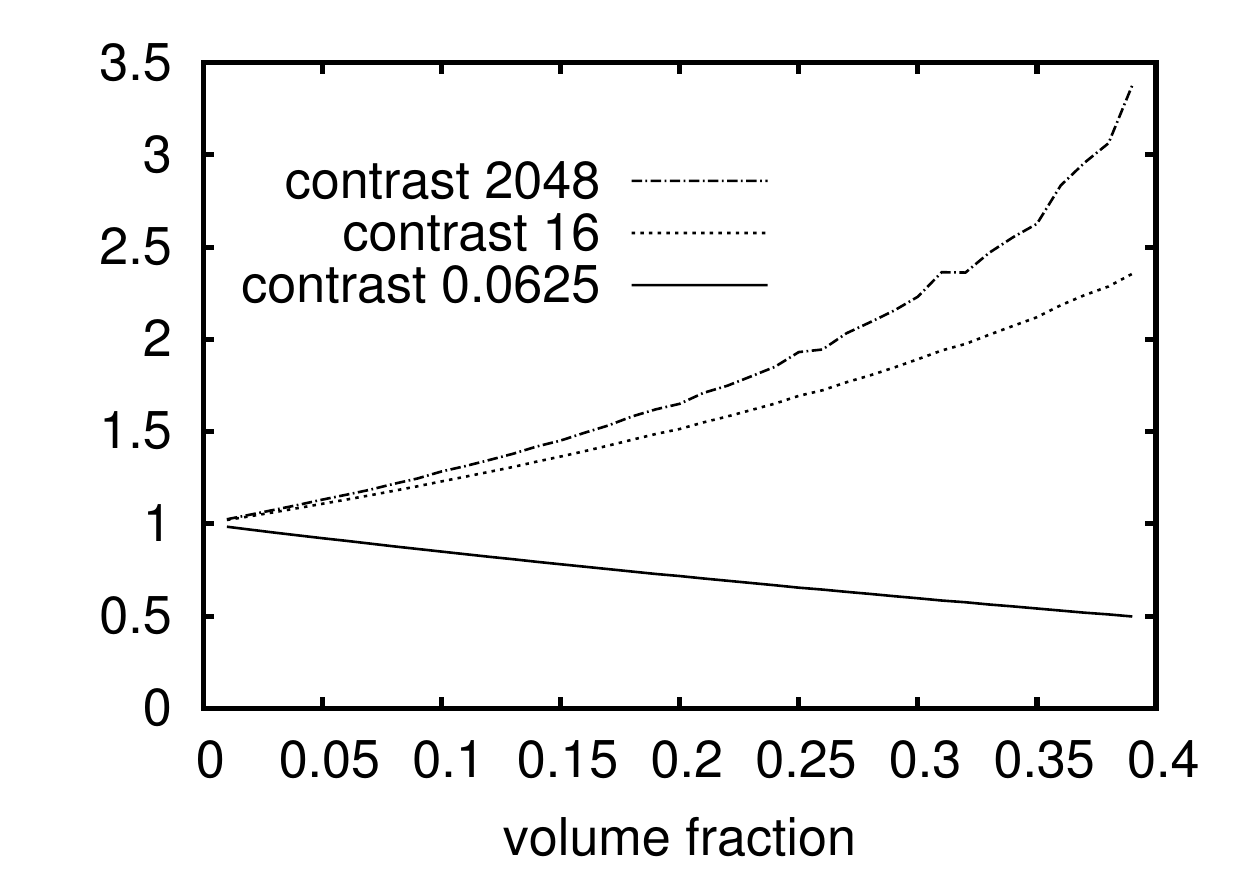} 
}
\subfigure[\, Shear modulus]{ 
   \includegraphics[width=0.45\linewidth]{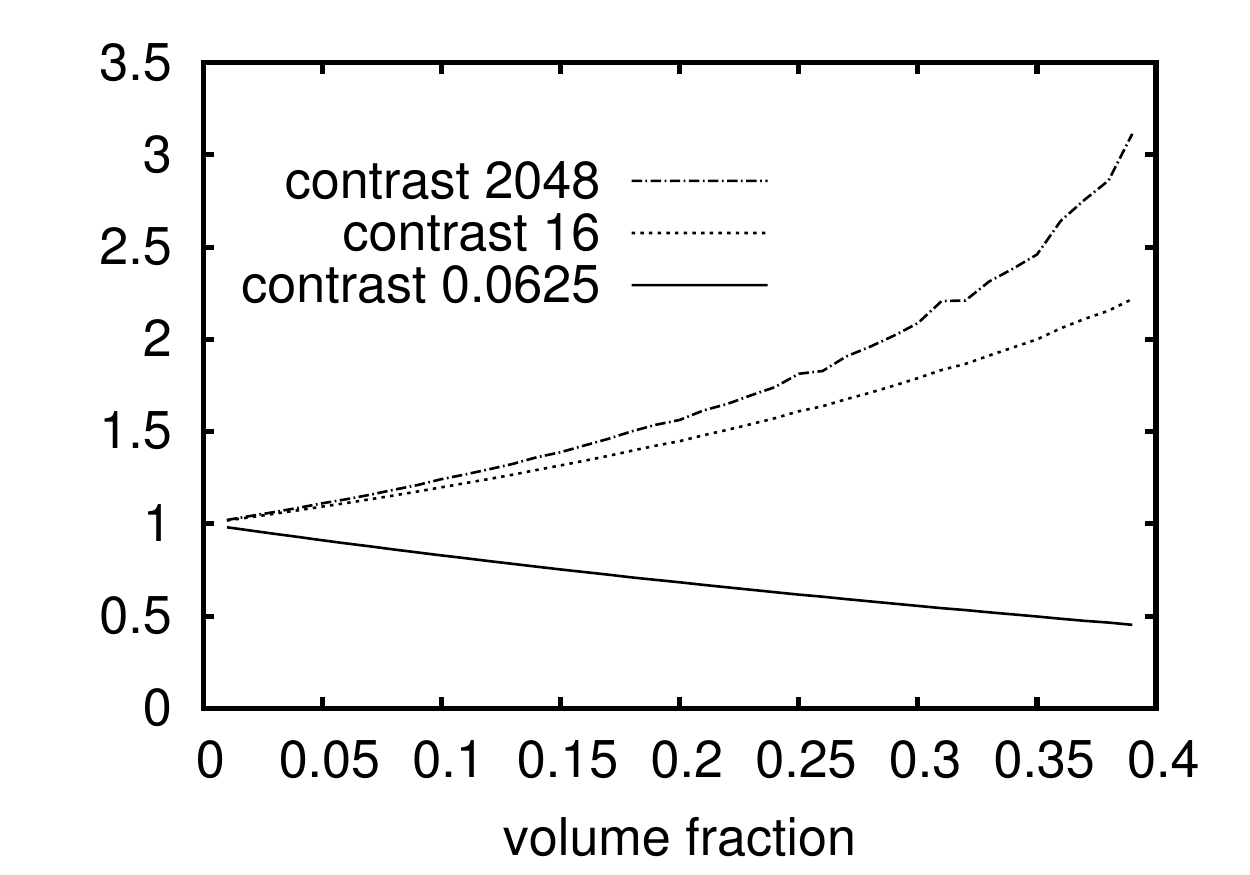}
 }
\caption{\label{fig:sp} The dependence of mechanical parameters of a composite
on the volume fraction of spherical inclusions. $n_{sp} =  20$, a similar picture for 
15 and 25.
} 
 \end{figure}

\begin{figure}[ht]  
\centering
\subfigure[\, Contrast 0.0625, normalized bulk modulus]{    
 \includegraphics[width=0.45\linewidth]{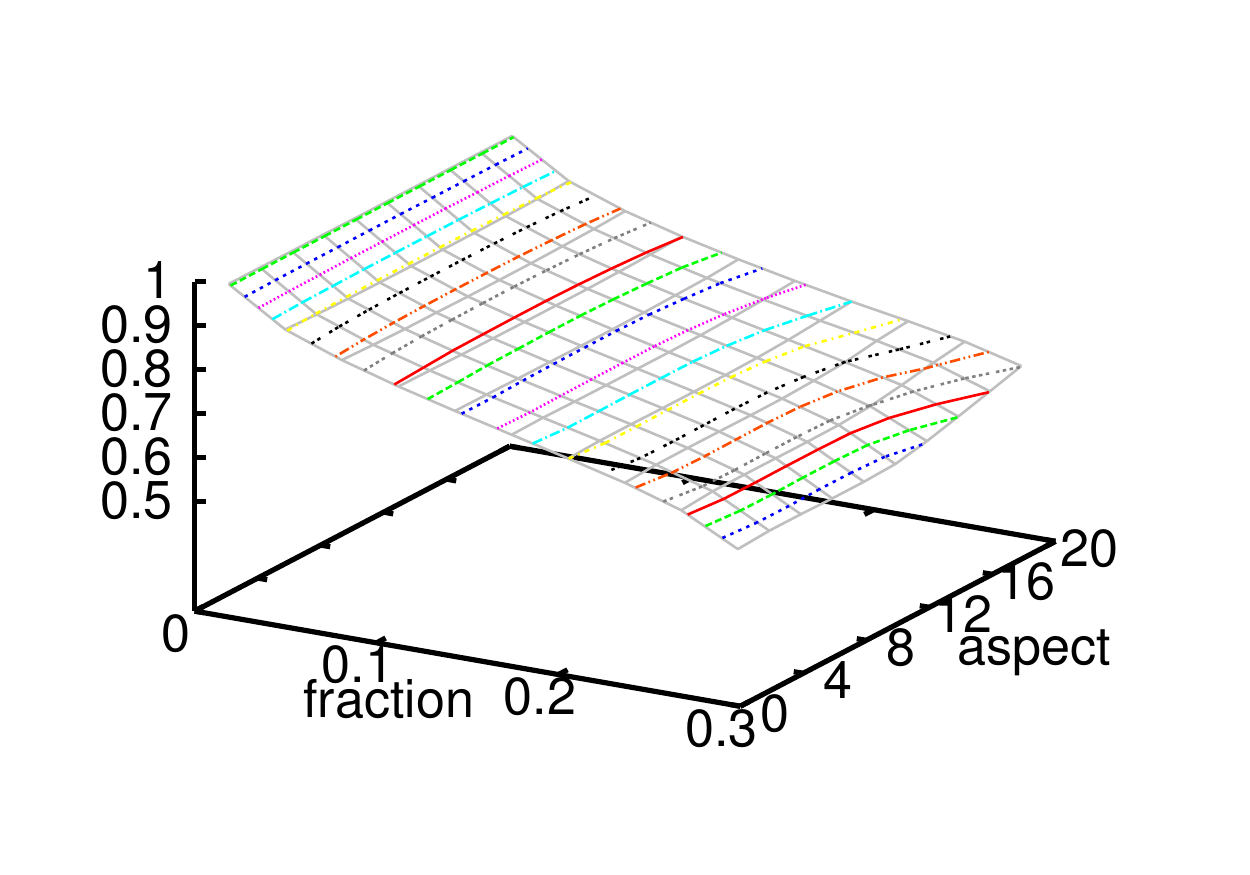} 
}
\subfigure[\, Contrast 16, normalized bulk modulus]{    
 \includegraphics[width=0.45\linewidth]{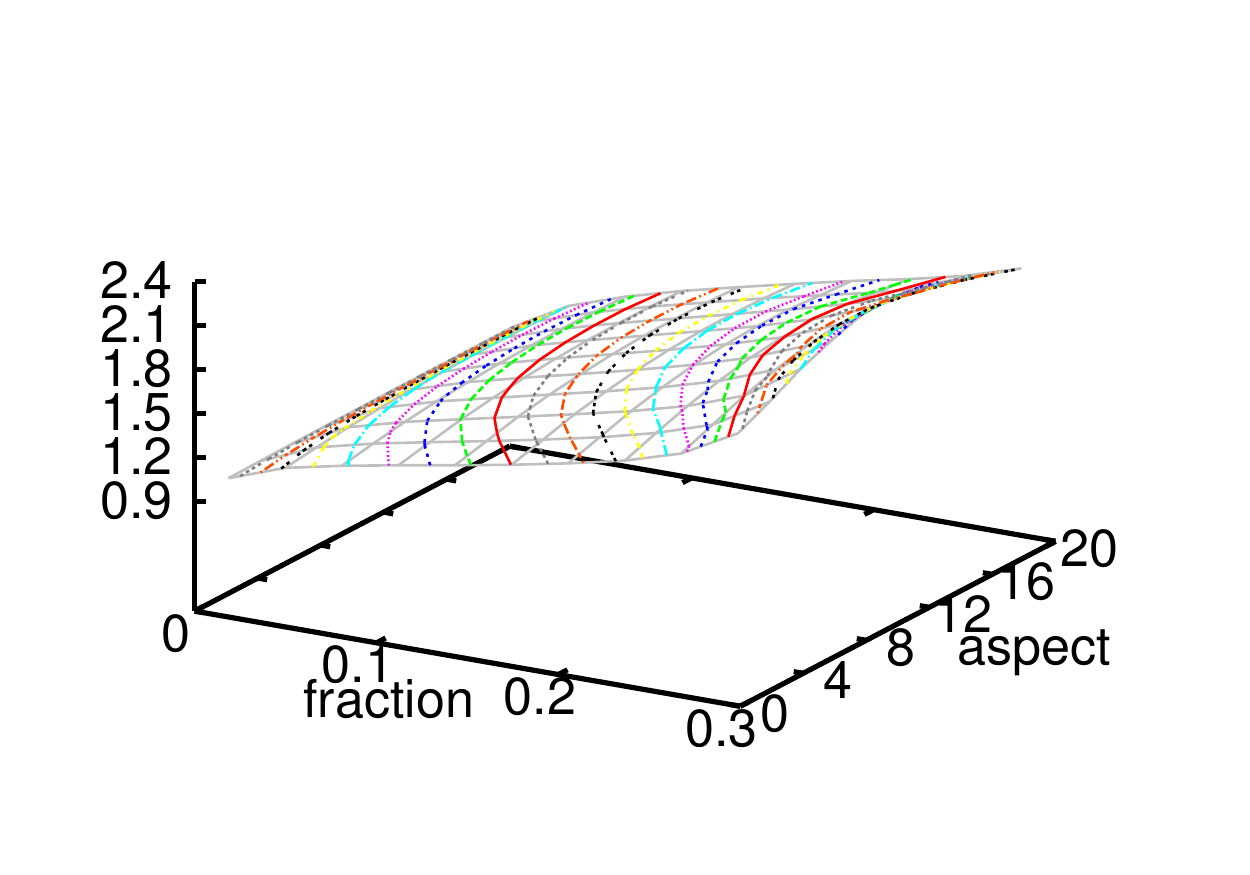} 
}
\subfigure[\, Contrast 2048, normalized bulk modulus]{    
 \includegraphics[width=0.45\linewidth]{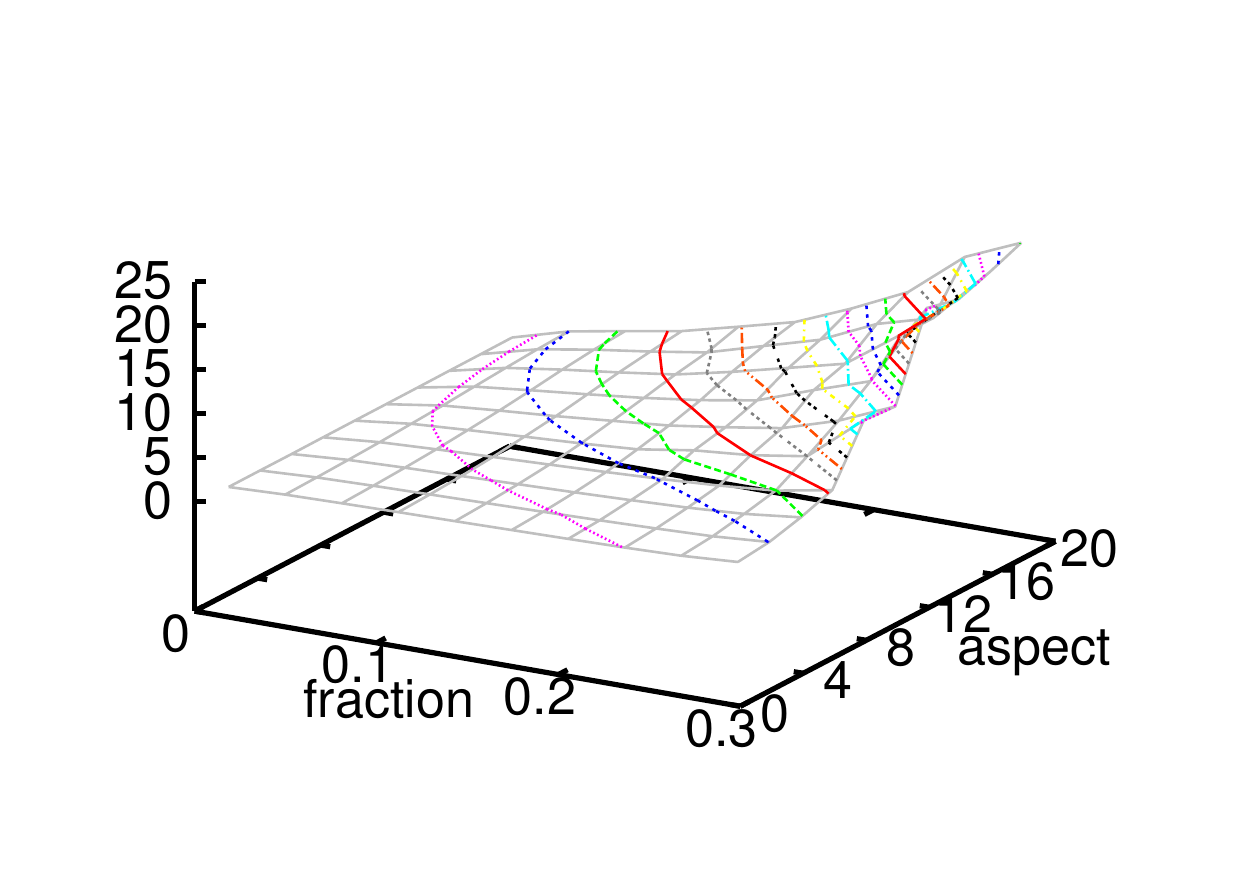} 
}
\subfigure[\, Contrast 2048, normalized shear modulus]{    
 \includegraphics[width=0.45\linewidth]{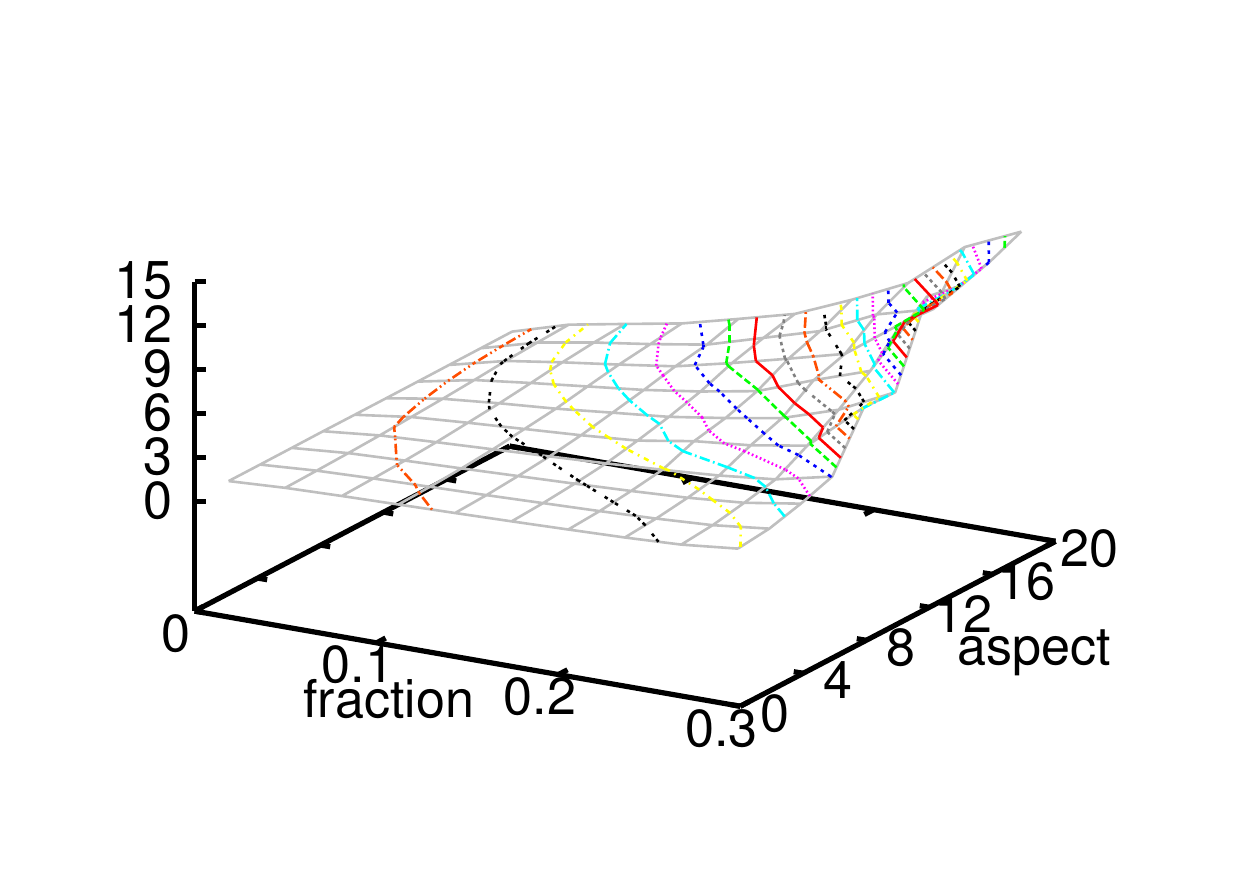} 
}

\caption{\label{fig:cyl} The dependence of mechanical parameters of a composite
on the volume fraction and aspect ratio of cylindrical inclusions.
$n_{cyl} = 20$
} 
 \end{figure}

To start with, let us consider only spherical inclusions (figure \ref{fig:sp})
and only cylindrical ones (figure \ref{fig:cyl}). 
For the spheres at contrast greater than $1$ the effective parameters 
increase non-linearly with the volume fraction, and decrease for the contrast less than 
$1$. The further the contrast is from $1$ the more this effect is pronounced.
The similar effect is observed for the cylinders, and in addition to this the reinforcement 
by longer  inclusions (with higher aspect ratio) is more efficient.

These first tests should be considered more as preparatory results and validation
check for the methods, since their outcome is rather predictable. In what follows we 
discuss more interesting tests.

\subsection{Advanced morphology analysis}
Let us now turn to more subtle questions related to analysis of the influence 
of morphology on the effective properties of composite materials, namely 
let us consider the composites reinforced by the mixture of globular and fiber-type 
inclusions and understand what factors in repartition of the inclusions to 
these two types influence the result.

\begin{figure}[ht]  
\centering
\subfigure[\, Contrast 0.0625, normalized bulk modulus]{    
 \includegraphics[width=0.45\linewidth]{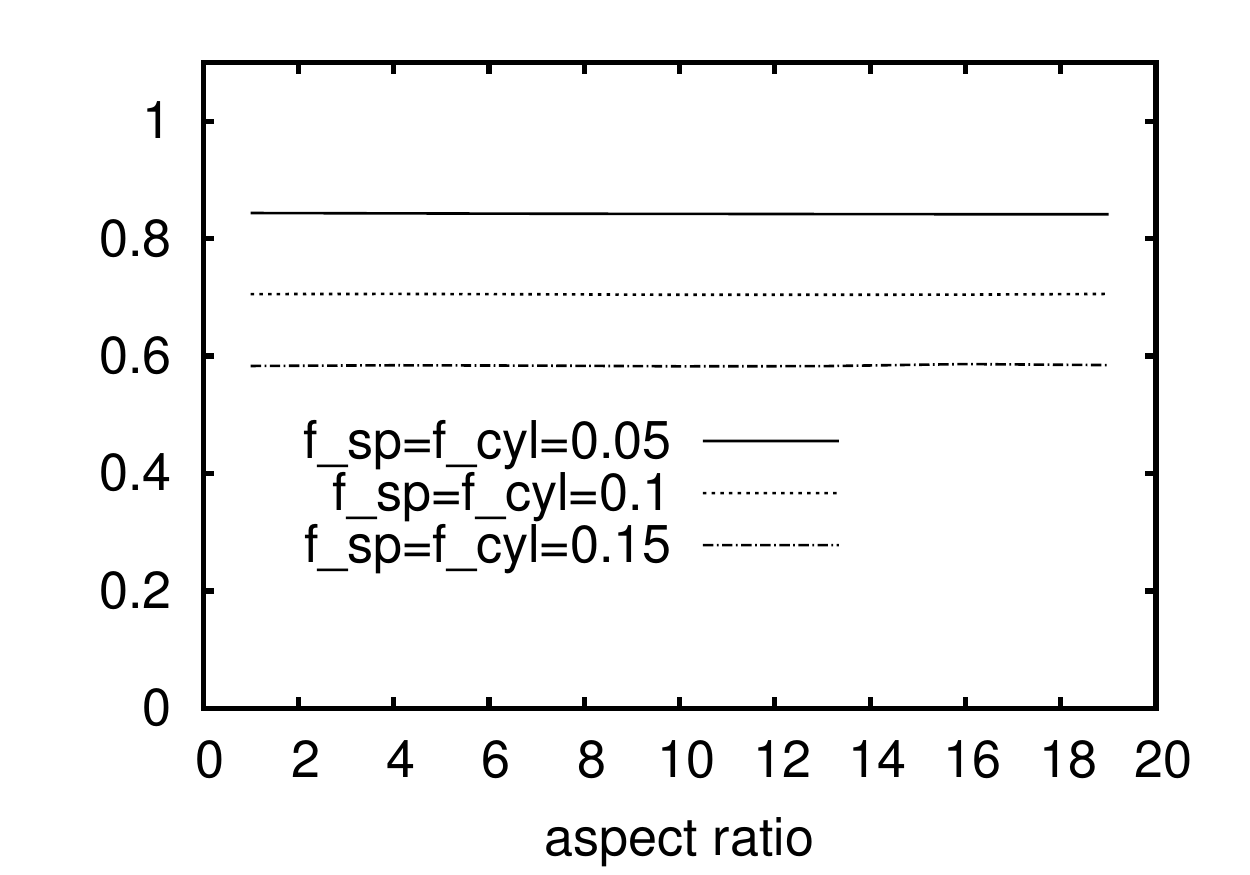} 
}
\subfigure[\, Contrast 16, normalized bulk modulus]{    
 \includegraphics[width=0.45\linewidth]{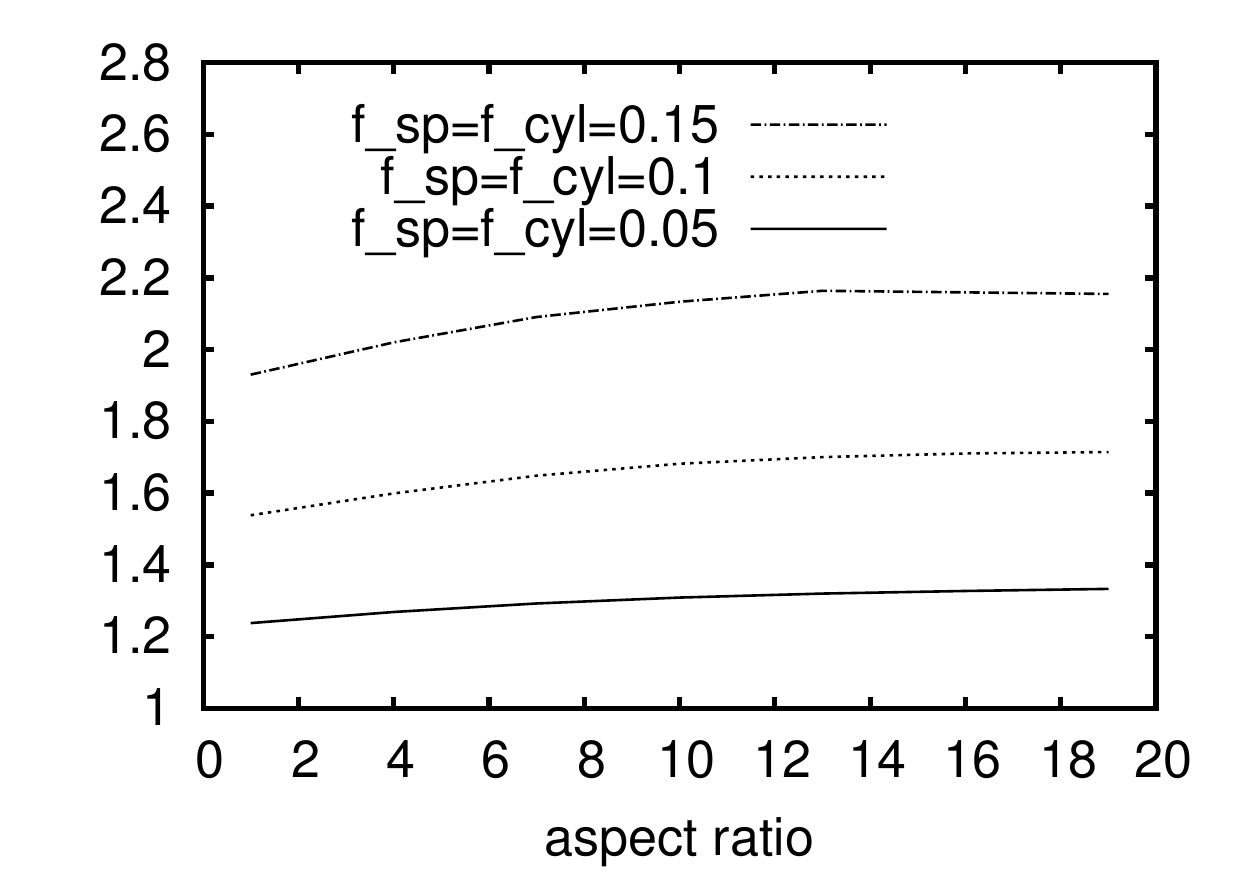} 
}
\subfigure[\, Contrast 256, normalized bulk modulus]{    
 \includegraphics[width=0.45\linewidth]{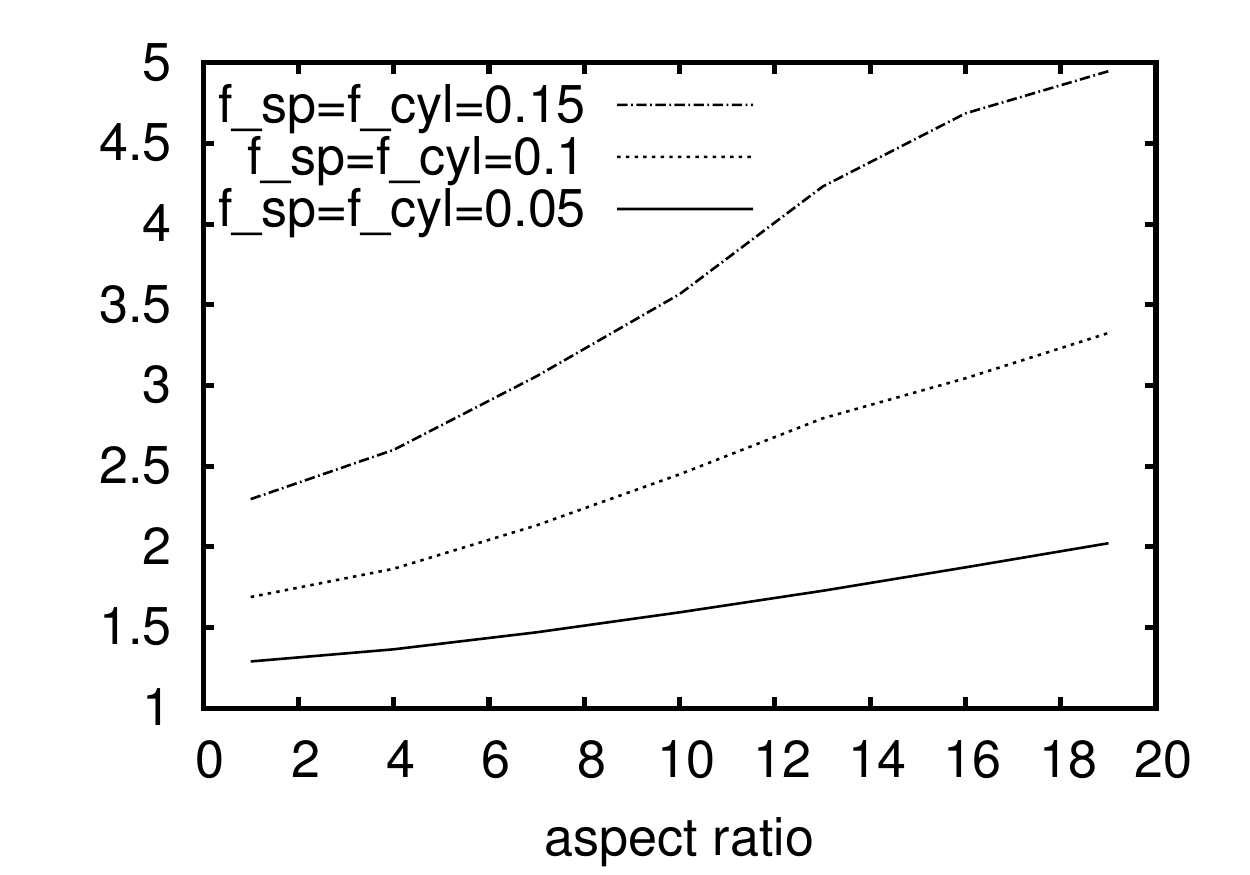} 
}
\subfigure[\, Contrast 2048, normalized bulk modulus]{    
 \includegraphics[width=0.45\linewidth]{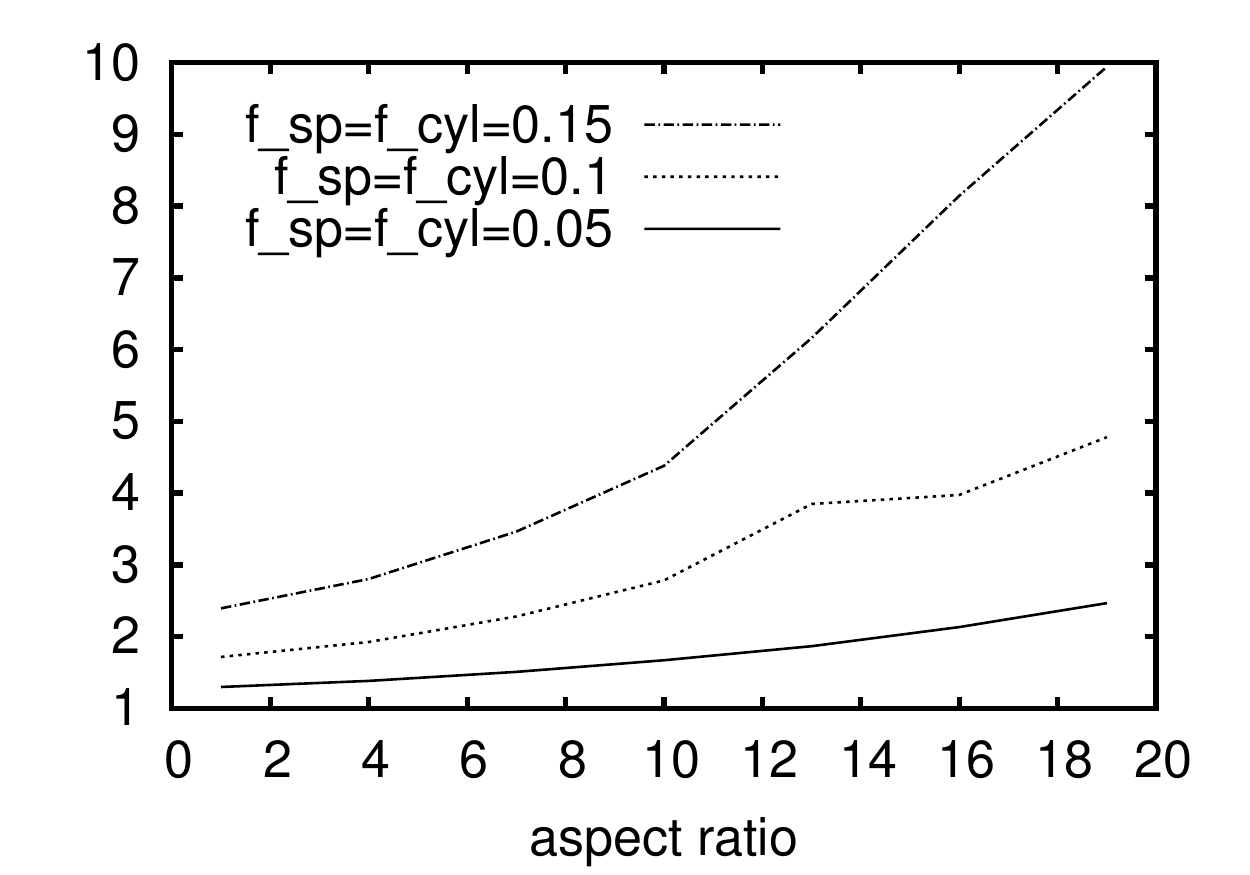} 
}

\caption{\label{fig:aspect} The dependence of mechanical parameters of a composite
material on the aspect ratio of the cylinders in the mixture of inclusions. Comparison for 
$n_{sp} = n_{cyl} = 20$.
}
 \end{figure}

In the first series of tests here we fix the overall number of inclusions of both type 
and study the dependence of the effective properties on the aspect ratio of the 
cylindrical inclusions for various volume fractions. The results (figure \ref{fig:aspect})
clearly show that for contrast greater than $1$ the composite is better reinforced 
with longer cylinders. The effect is certainly better visible for higher contrast.
Although as expected the presence of spherical inclusions in the mixture 
does not permit to reach the same values of parameters of the homogenized medium 
as in the case of the same volume fraction formed by cylinders only (cf. figure \ref{fig:cyl}).
The opposite effect is present for values of contrast less than $1$, but quantitatively it is 
less pronounced.

\begin{figure}[ht]  
\centering
\subfigure[\, $f_{sp}\!=\!f_{cyl} = 0.05$, contrast 0.0625, normalized bulk modulus]{    
 \includegraphics[width=0.43\linewidth]{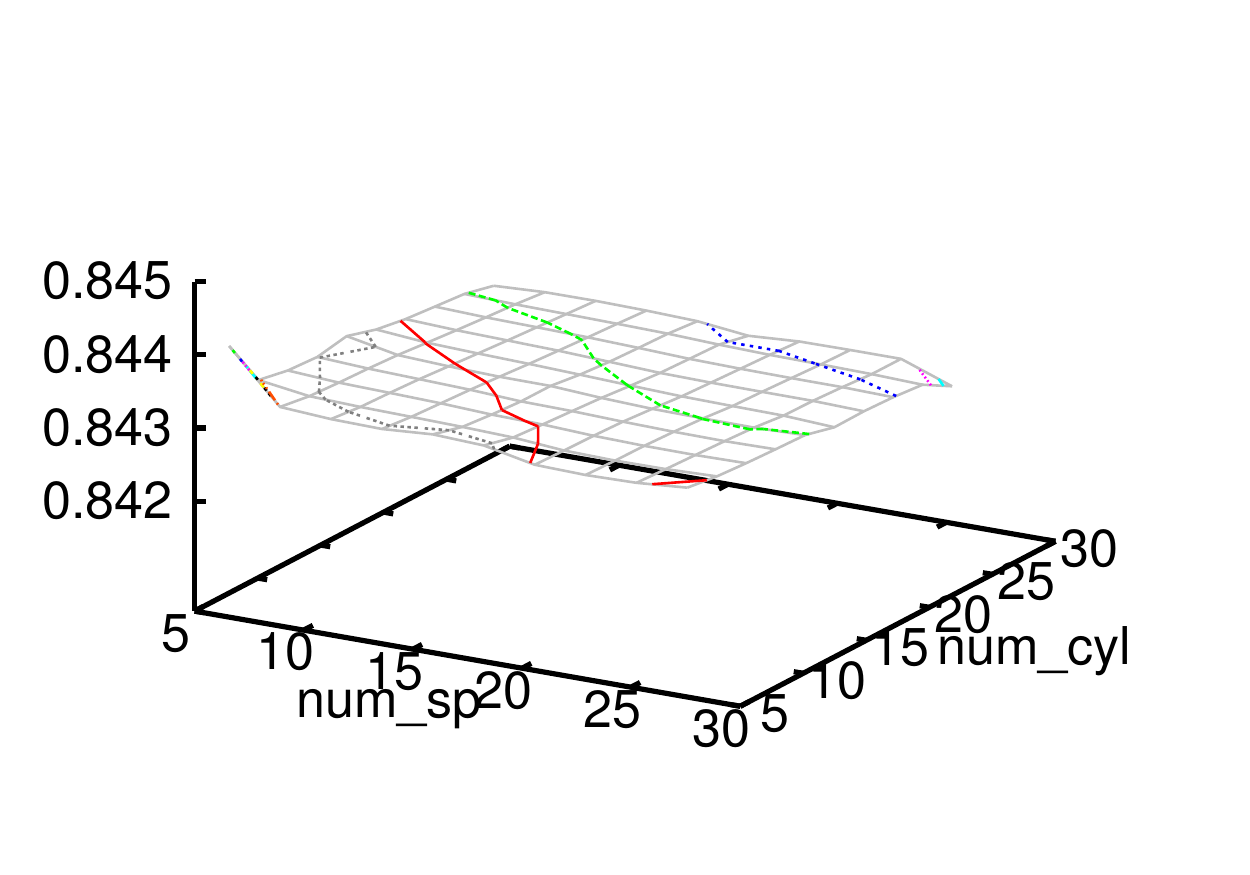} 
}
\subfigure[\, $f_{sp} = f_{cyl} = 0.05$, contrast 2048, normalized bulk modulus]{    
 \includegraphics[width=0.43\linewidth]{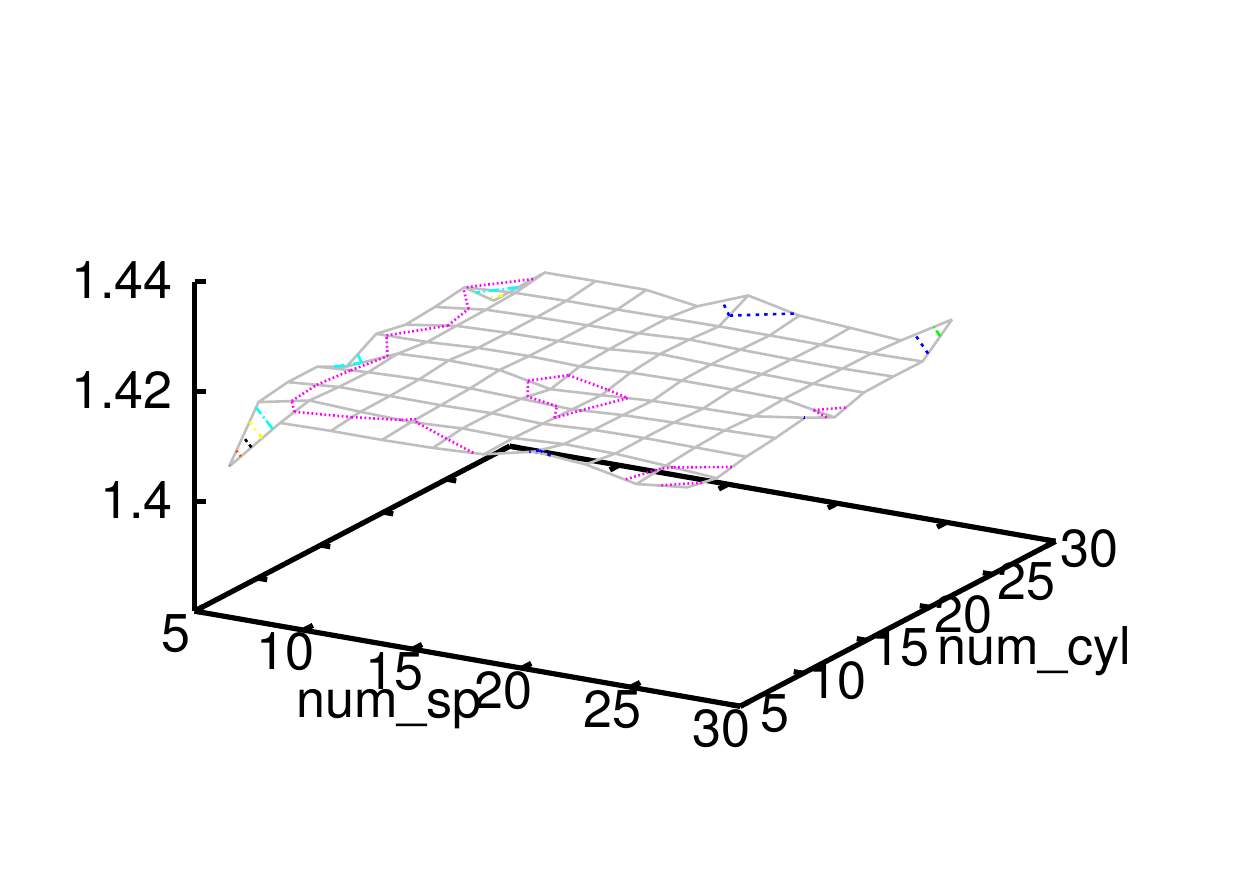} 
}
\subfigure[\, $f_{sp} = f_{cyl} = 0.1$, contrast 0.0625 normalized bulk modulus]{    
 \includegraphics[width=0.43\linewidth]{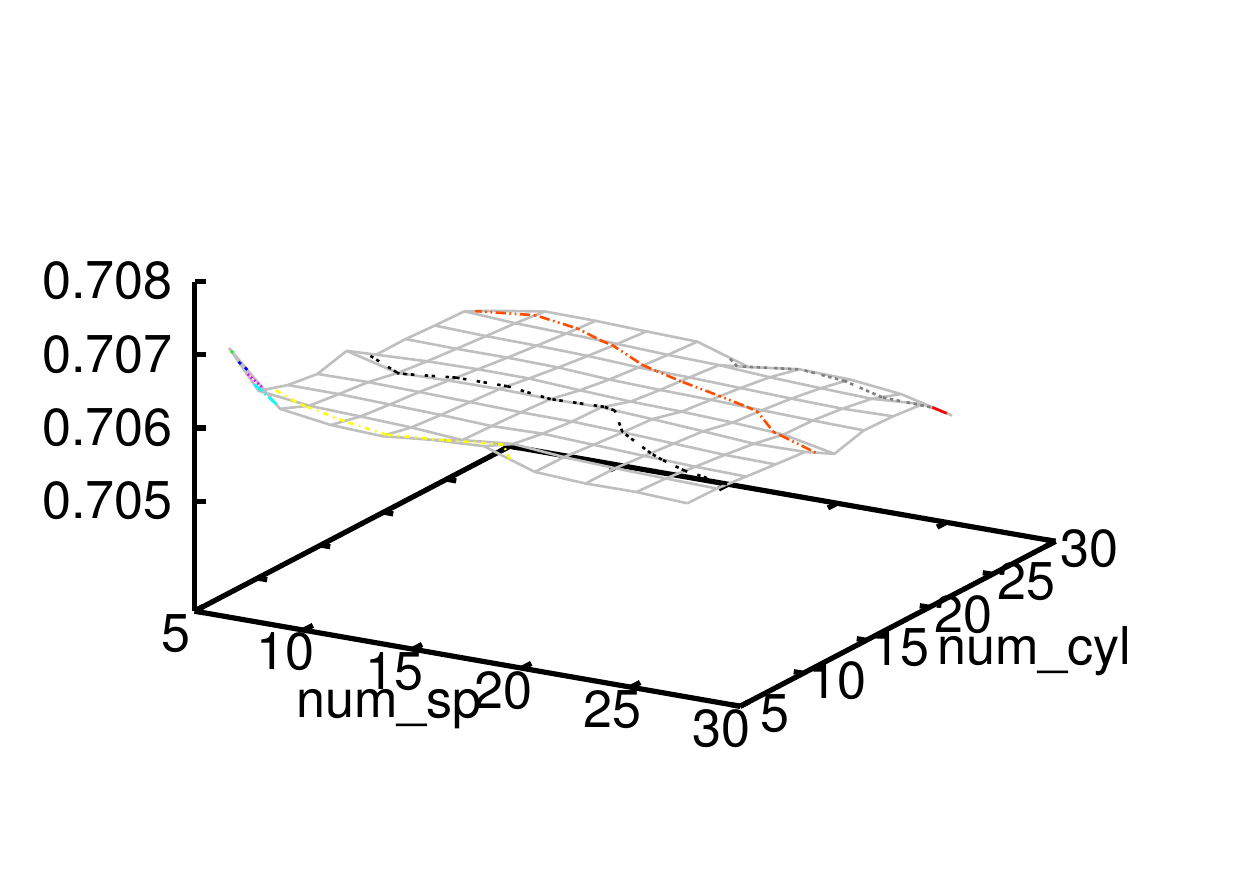} 
}
\subfigure[\, $f_{sp} = f_{cyl} = 0.1$, contrast 2048 normalized bulk modulus]{    
 \includegraphics[width=0.43\linewidth]{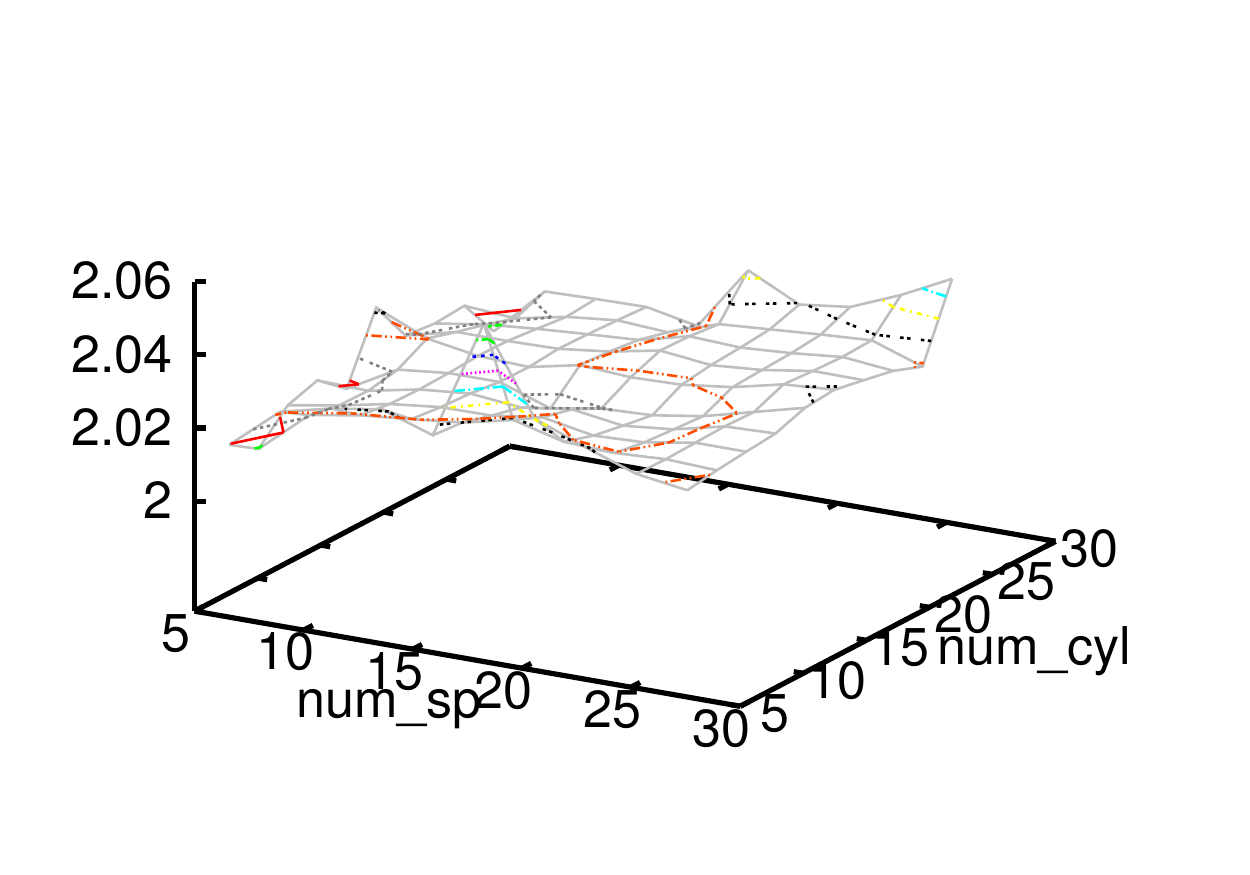} 
}
\subfigure[\, $f_{sp} = f_{cyl} = 0.15$, contrast 0.0625 normalized bulk modulus]{    
 \includegraphics[width=0.43\linewidth]{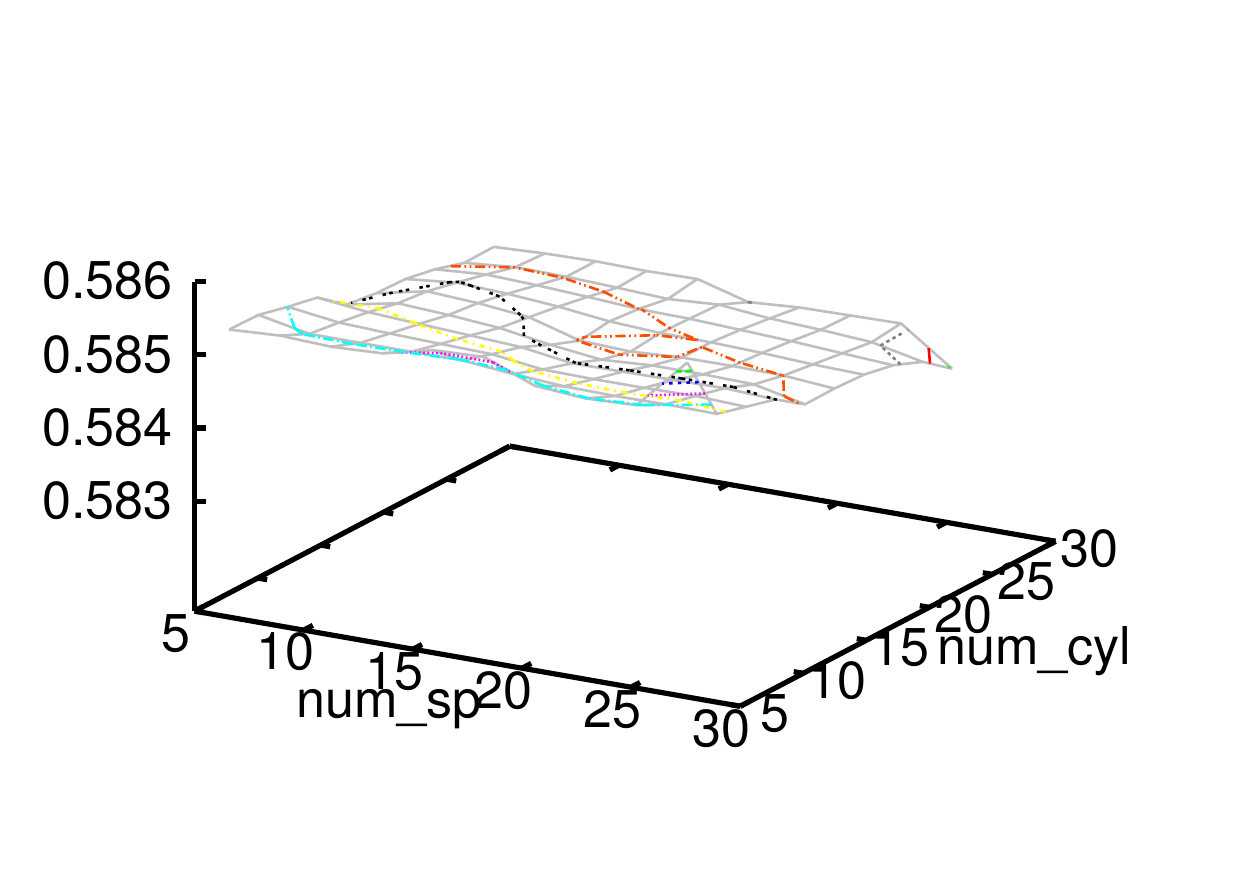} 
}
\subfigure[\, $f_{sp} = f_{cyl} = 0.15$, contrast 2048 normalized bulk modulus]{    
 \includegraphics[width=0.43\linewidth]{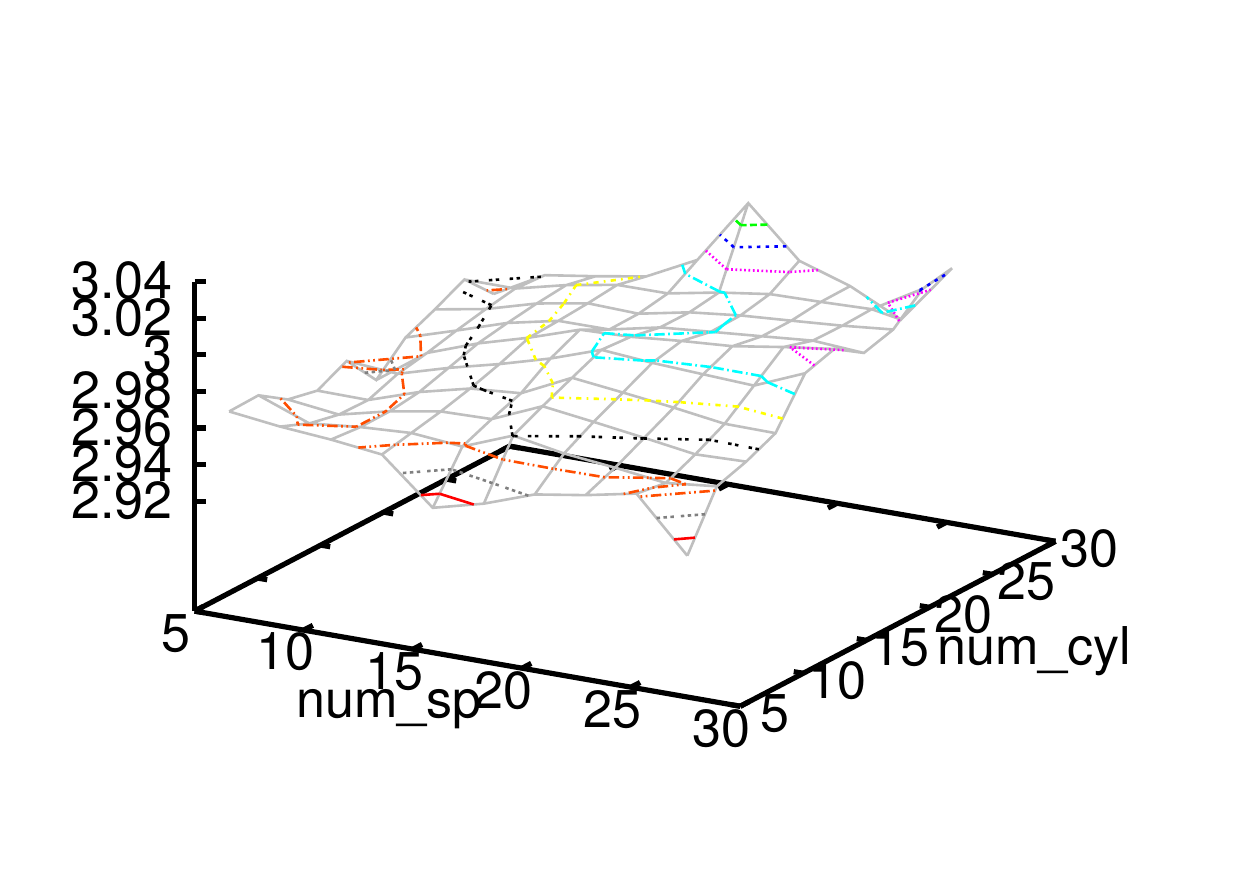} 
}

\caption{\label{fig:num} The dependence of mechanical parameters of a composite material
on the number of various inclusions for fixed volume fraction.
Note that the scale on these plots is different from all the others: the difference between 
the minimal and the maximal values is almost absorbed by the confidence intervals, except the 
corner with the small number of inclusions (cf. also figure \ref{fig:rve}).
} 
 \end{figure}

 Let us now fix the volume fraction of each type of inclusions and vary the number of them.
 The figure \ref{fig:num} shows that the reinforcement/weakening is slightly more efficient 
 with a large number of cylinders. Although looking at numerical values one sees that 
 the effect is rather subtle and can be neglected in the global analysis. 
 Notice that there is a saturation phenomenon when the total number of inclusions is small,
 which is in perfect agreement with the above discussion about the size of an RVE and figure 
 \ref{fig:rve}.

\begin{figure}[ht]  
\centering
\subfigure[\, contrast 0.0625, normalized bulk modulus]{    
 \includegraphics[width=0.45\linewidth]{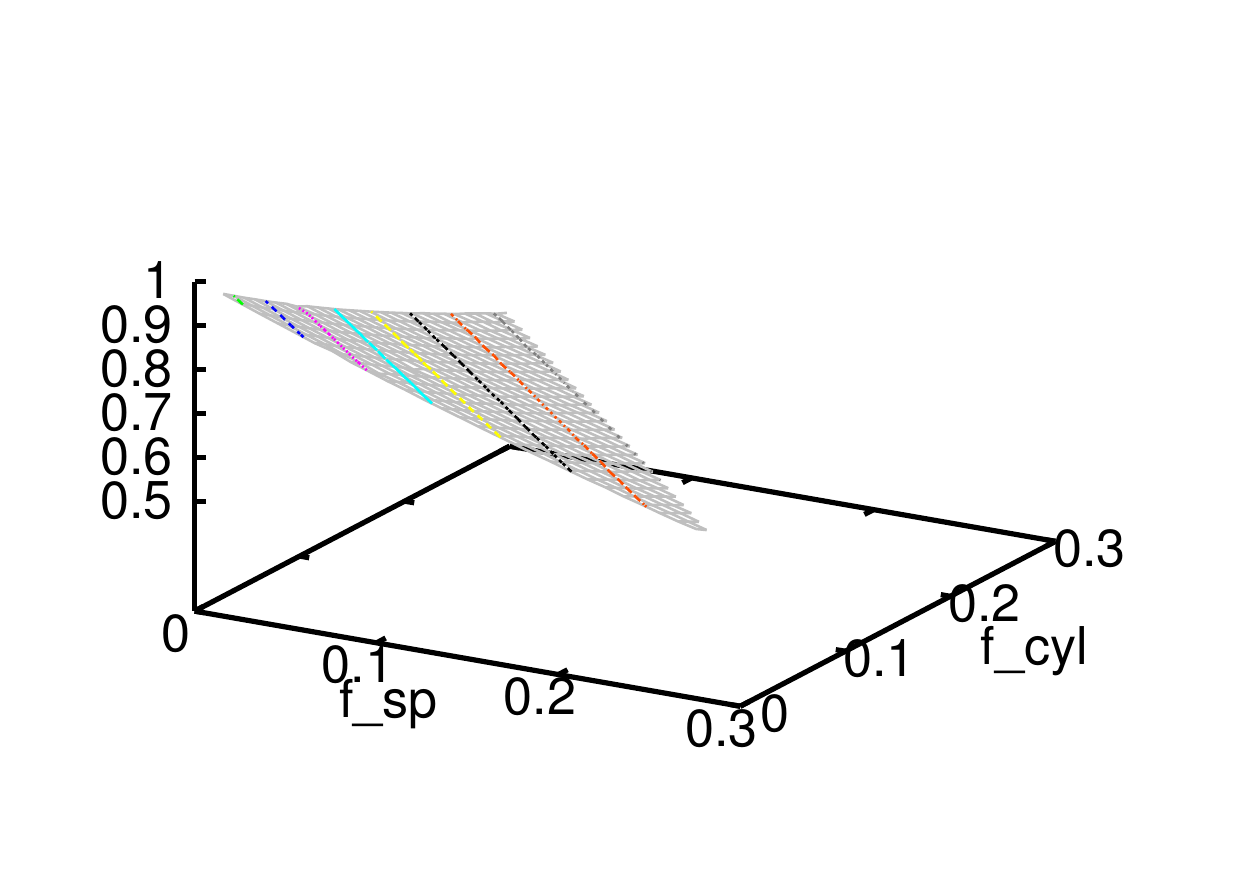} 
}
\subfigure[\, contrast 0.0625, normalized bulk modulus]{    
 \includegraphics[width=0.45\linewidth]{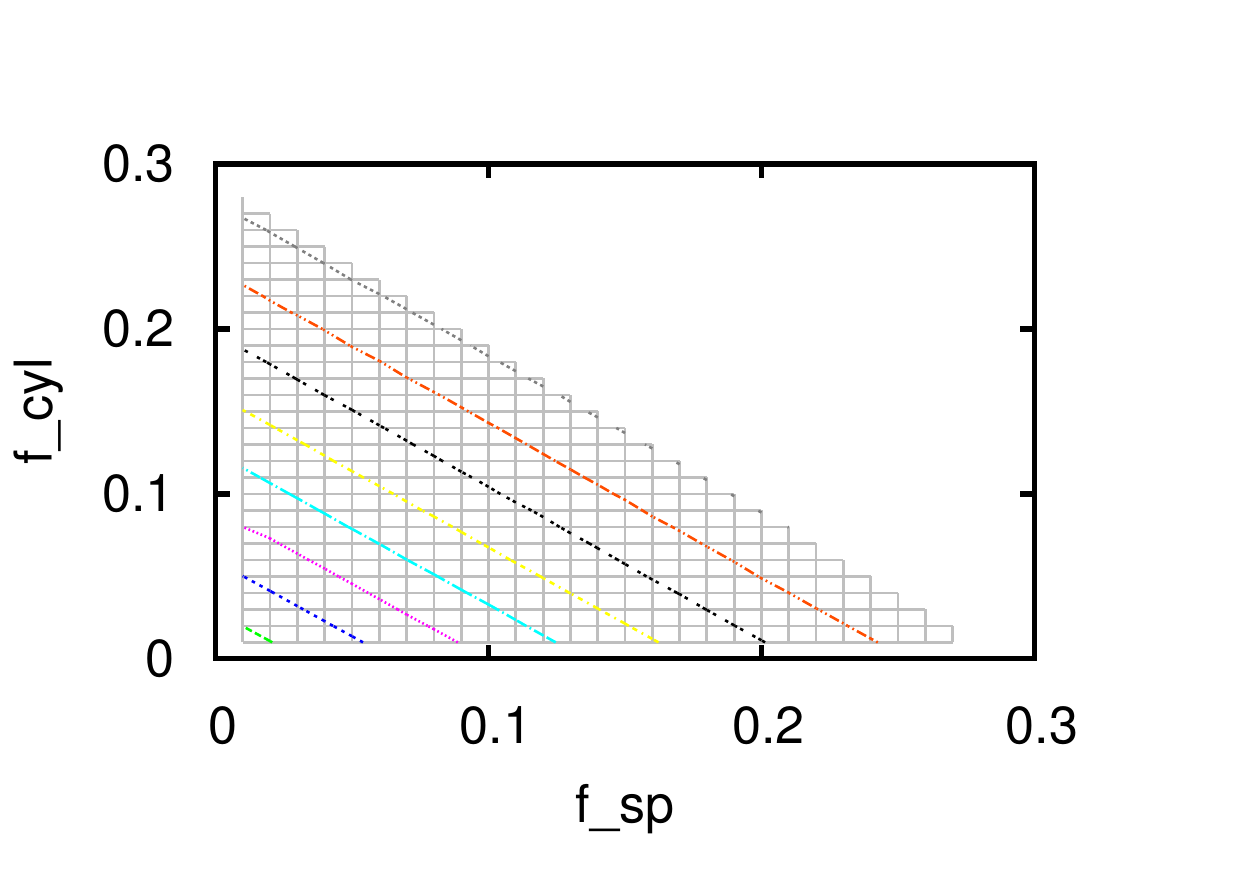} 
}

\subfigure[\, contrast 16, normalized bulk modulus]{    
 \includegraphics[width=0.45\linewidth]{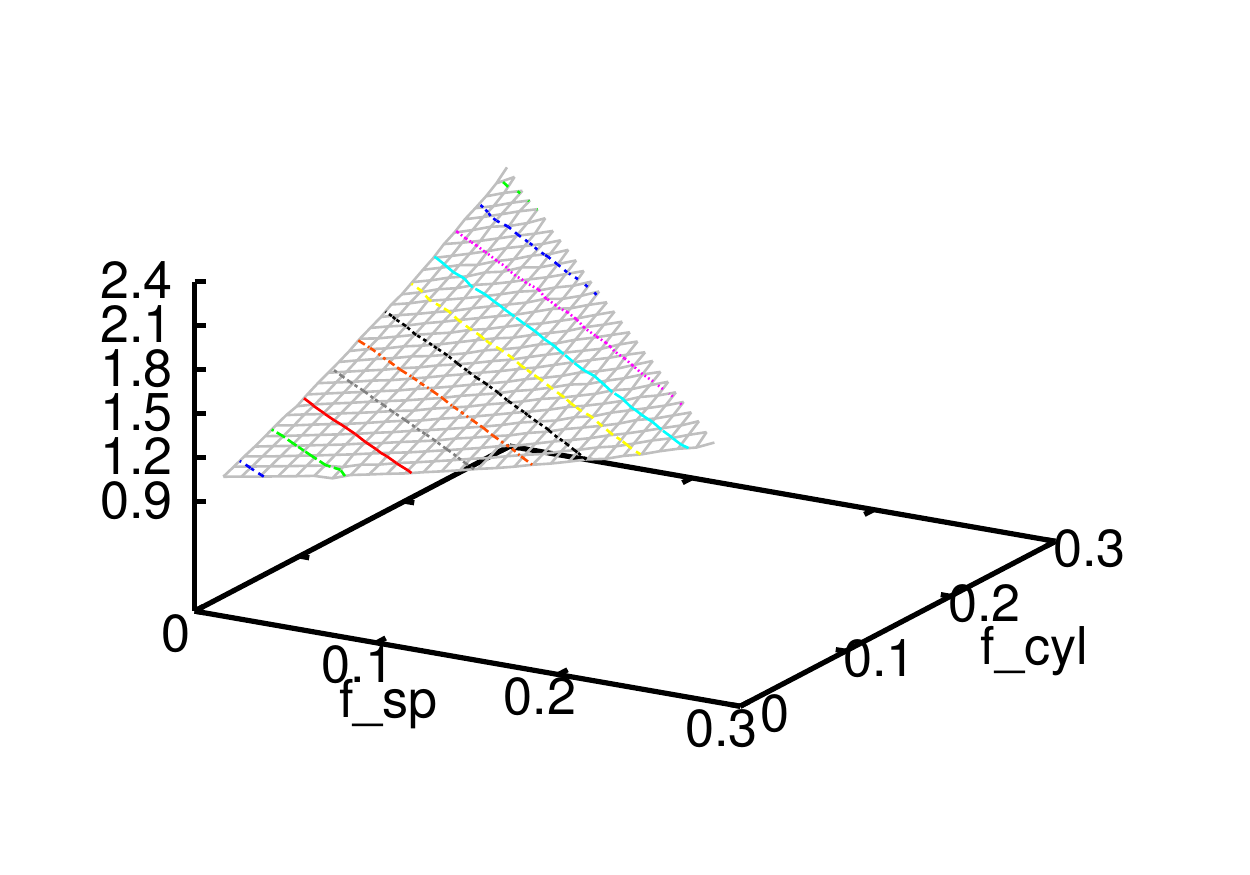} 
}
\subfigure[\, contrast 16, normalized bulk modulus]{    
 \includegraphics[width=0.45\linewidth]{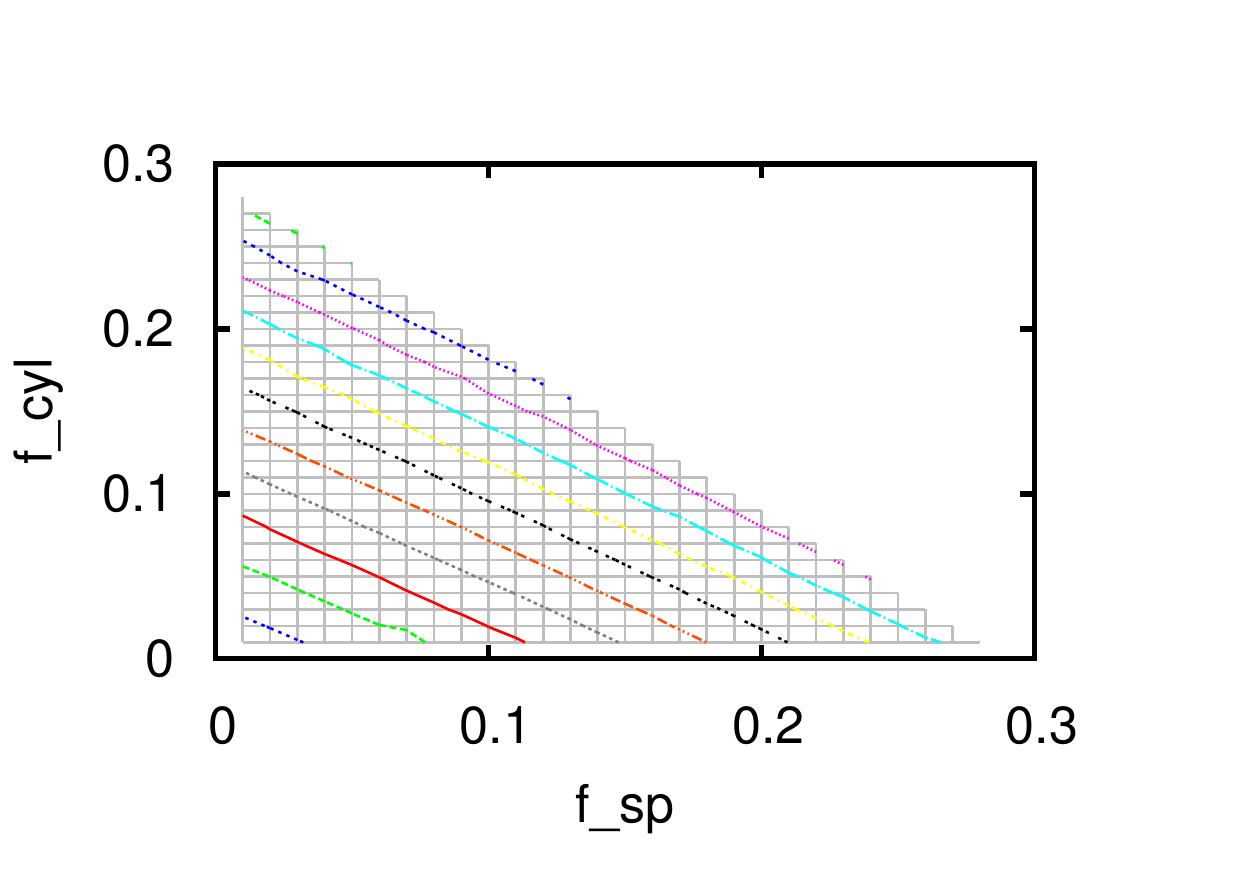} 
}
\subfigure[\, contrast 2048, normalized bulk modulus]{    
 \includegraphics[width=0.45\linewidth]{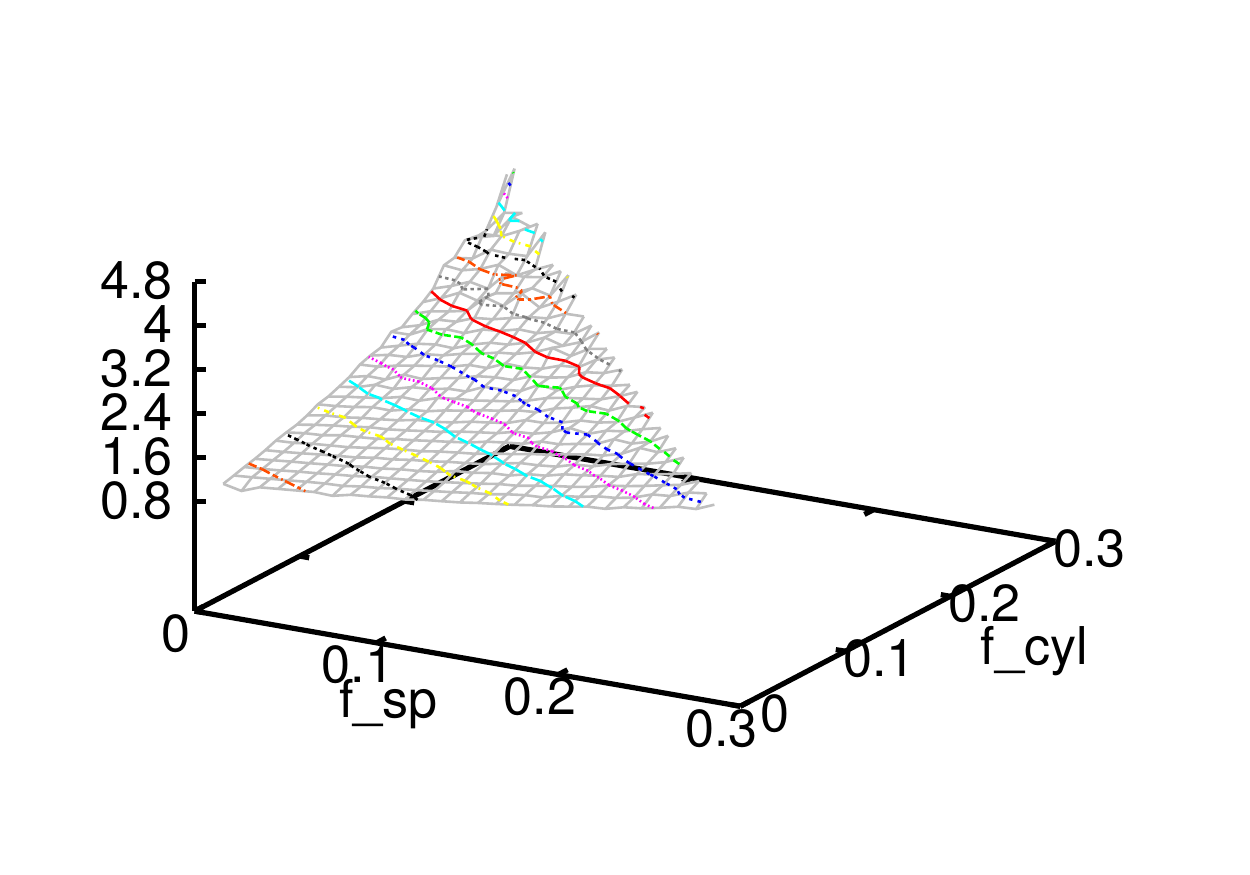} 
}
\subfigure[\, contrast 2048, normalized bulk modulus]{    
 \includegraphics[width=0.45\linewidth]{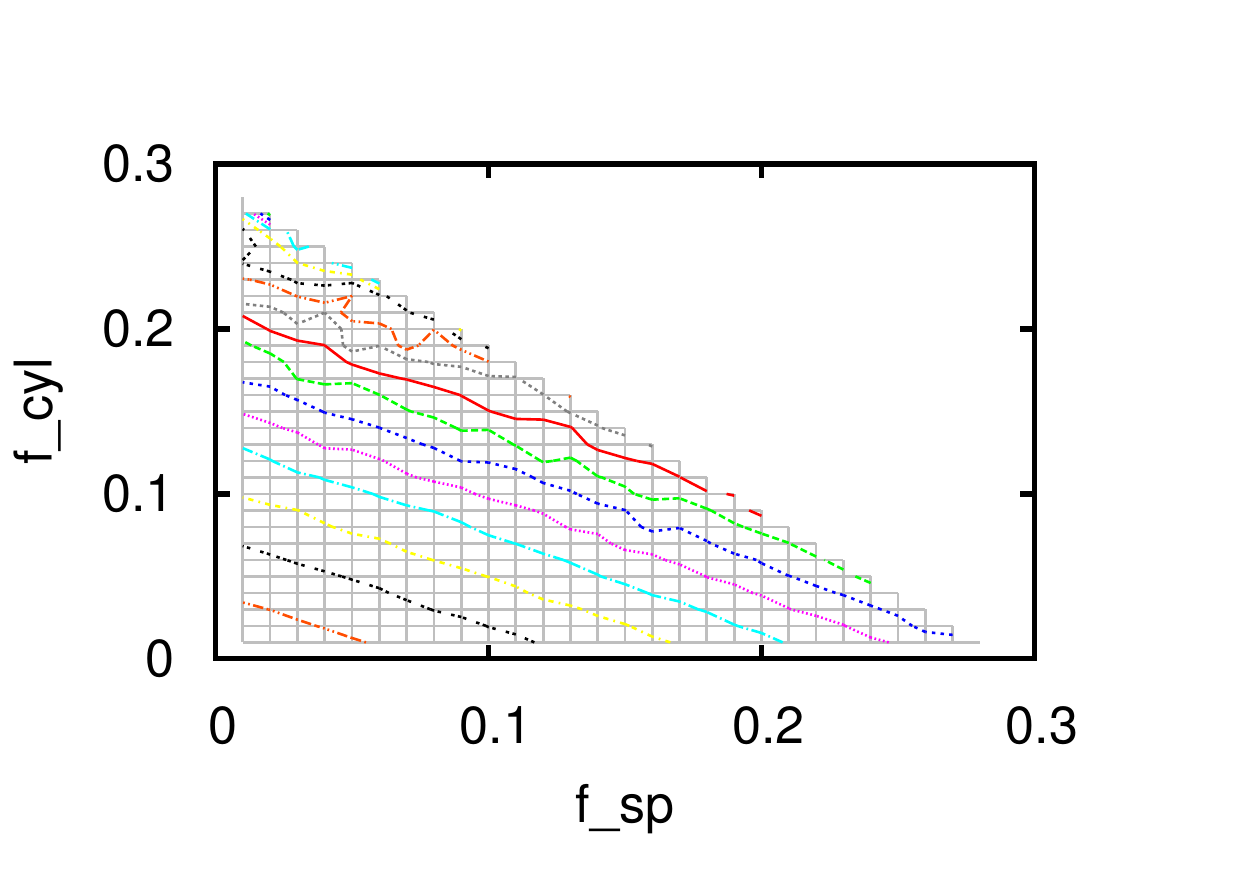} 
}

\caption{\label{fig:cyl_vs_sp} The dependence of mechanical parameters of a composite material
on the repartition of inclusions volume between spheres and cylinders. $n_{sp}=n_{cyl} = 20$.
3D plot on the left, level sets map on the right.
} 
 \end{figure}

The most interesting series of tests from this group is the study of the influence 
of repartition of the volume of inclusions between spheres and cylinders. 
The diagonals of each plot on the figure \ref{fig:cyl_vs_sp} represent 
the volume fractions of two types of inclusions with a fixed sum. 
One can again notice that the reinforcement/weakening effect is better observed 
for cylinders than for spheres.

 \clearpage

\subsection{Imperfections}
The following two series of tests represent the most true to life 
situations when the inclusions are not of ideal shape. 
With the algorithms from \cite{VDP} %, SVDP} 
we are able to generate various types 
of imperfections including perturbing the surfaces of inclusions by waves 
and taking out parts of the inclusions preserving the overall volume fraction
(figure \ref{fig:imperfections}). Let us mention that generating 
these imperfections we still suppose that the interface between the two materials 
is perfect, i.e. no voids or discontinuities are created. This 
proves to be a reasonable assumption for mechanical properties, 
let us note though that 
if one studies for example thermal or electrical conductivity 
the result is more subtle (cf. \cite{yvonnet}). %\cite{SVDP-therm, yvonnet}). 

\begin{figure}[ht]  
\centering
\subfigure[\, aspect 3, normalized bulk modulus]{    
 \includegraphics[width=0.45\linewidth]{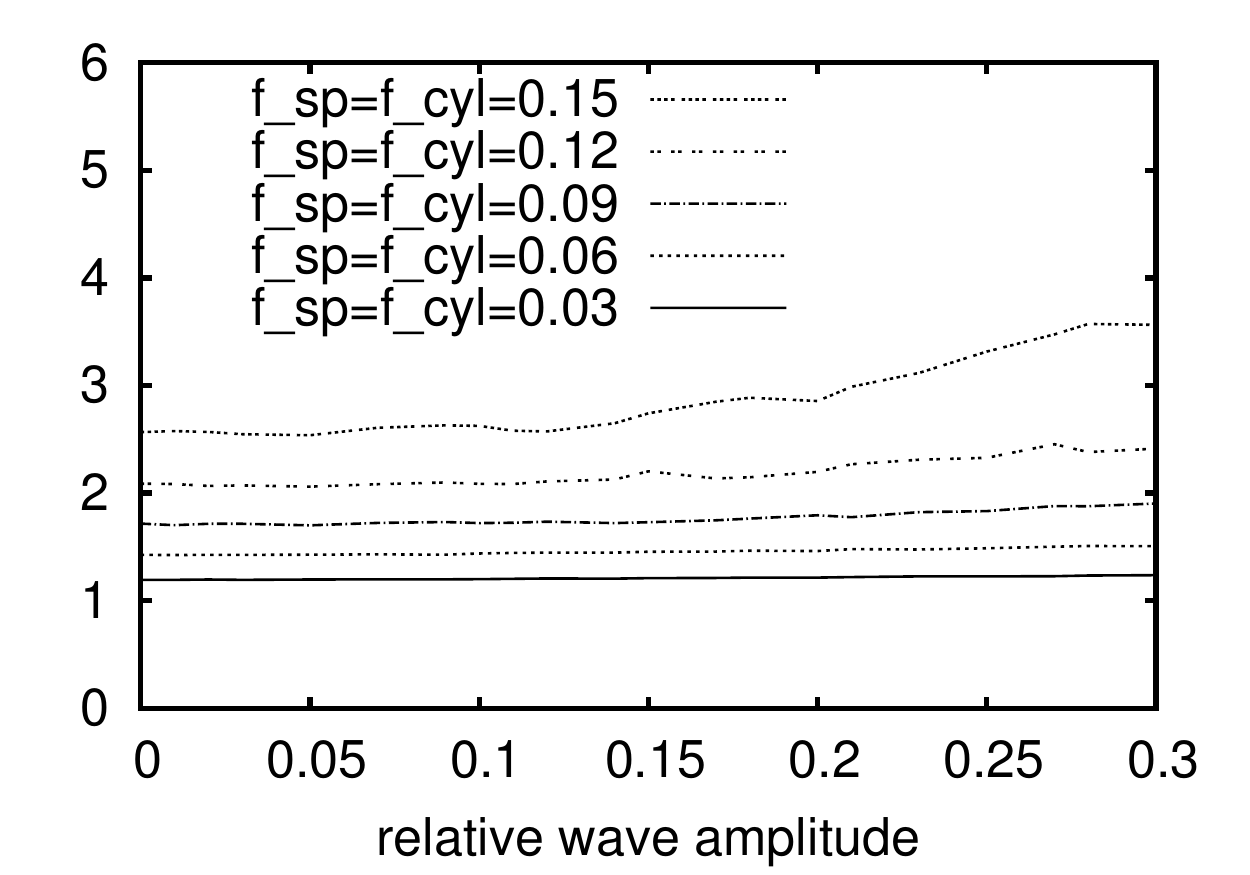} 
}
\subfigure[\, aspect 6, normalized bulk modulus]{    
 \includegraphics[width=0.45\linewidth]{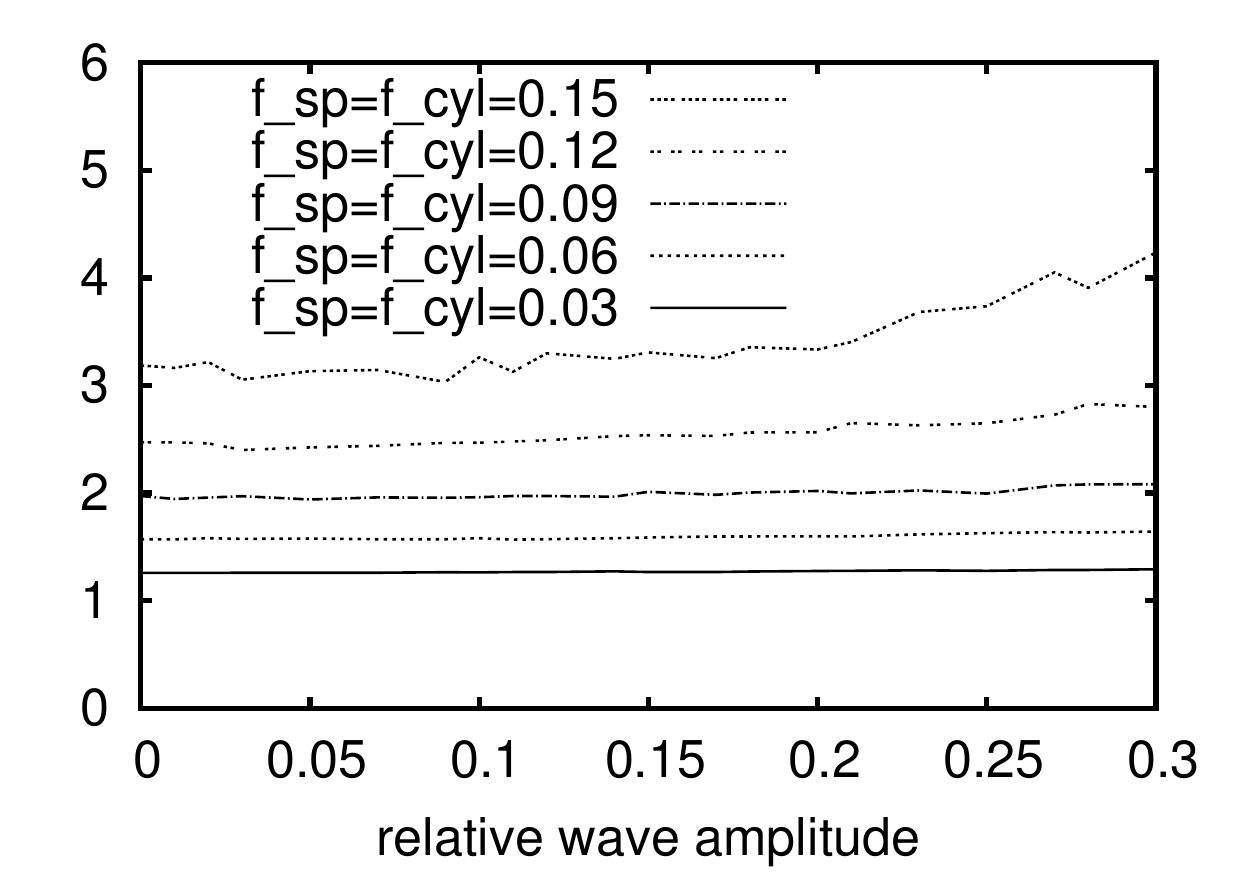} 
}

\subfigure[\, aspect 9, normalized bulk modulus]{    
 \includegraphics[width=0.45\linewidth]{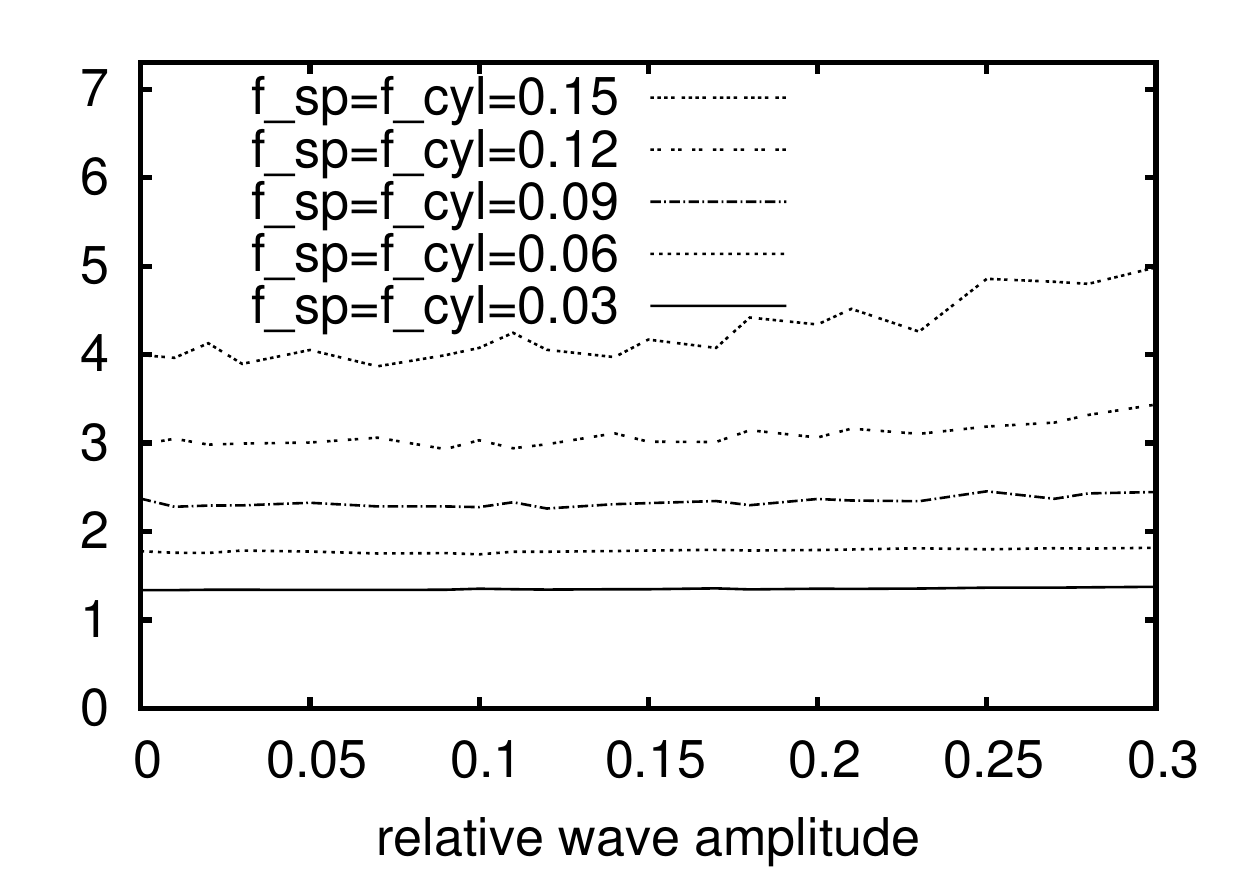} 
}
\subfigure[\, aspect 9, normalized shear modulus]{    
 \includegraphics[width=0.45\linewidth]{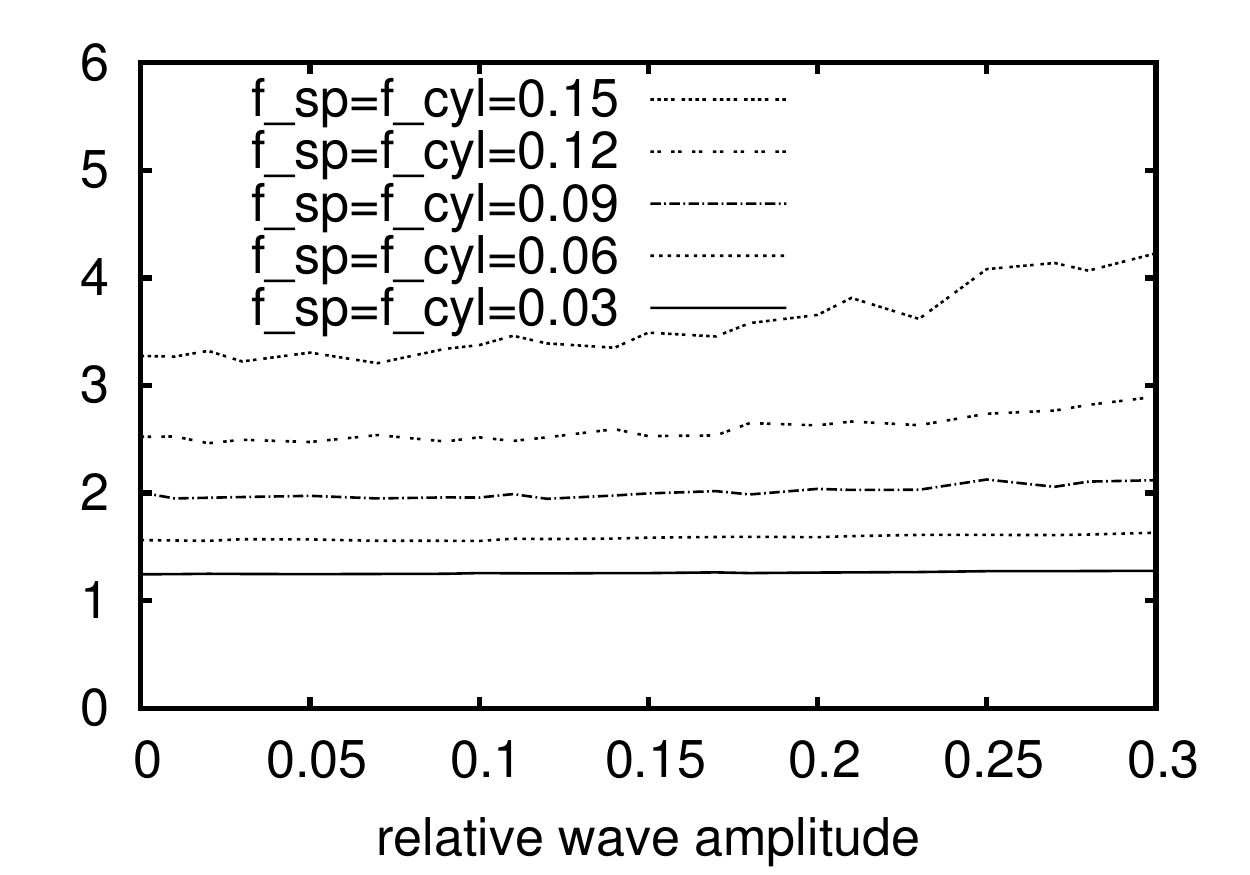} 
}

\caption{\label{fig:curved} The dependence of mechanical parameters of a composite
on the waving of the surface of inclusions. $n_{sp}=n_{cyl} = 20$, contrast 2048.
} 
 \end{figure}

In the first case the main parameter is the relative wave amplitude, i.e. the ratio
between the amplitude of perturbations of the surface and some characteristic size of 
the inclusions (radii of spheres and cylinders in the performed computations). 
A typical dependence of the effective properties on this parameter for several aspect 
ratios of cylinders in the mixture is presented on figure \ref{fig:curved}.
The trends observed there leave no doubt that such perturbations in fact contribute to 
more efficient reinforcement of the material. Our computations show that the effect is more pronounced when 
the composite is already well reinforced, i.e. at higher volume fraction, larger 
aspect ratio of cylinders, or significant contrast between two phases. 

\begin{figure}[ht]  
\centering
\subfigure[\, aspect 3, normalized bulk modulus]{    
 \includegraphics[width=0.45\linewidth]{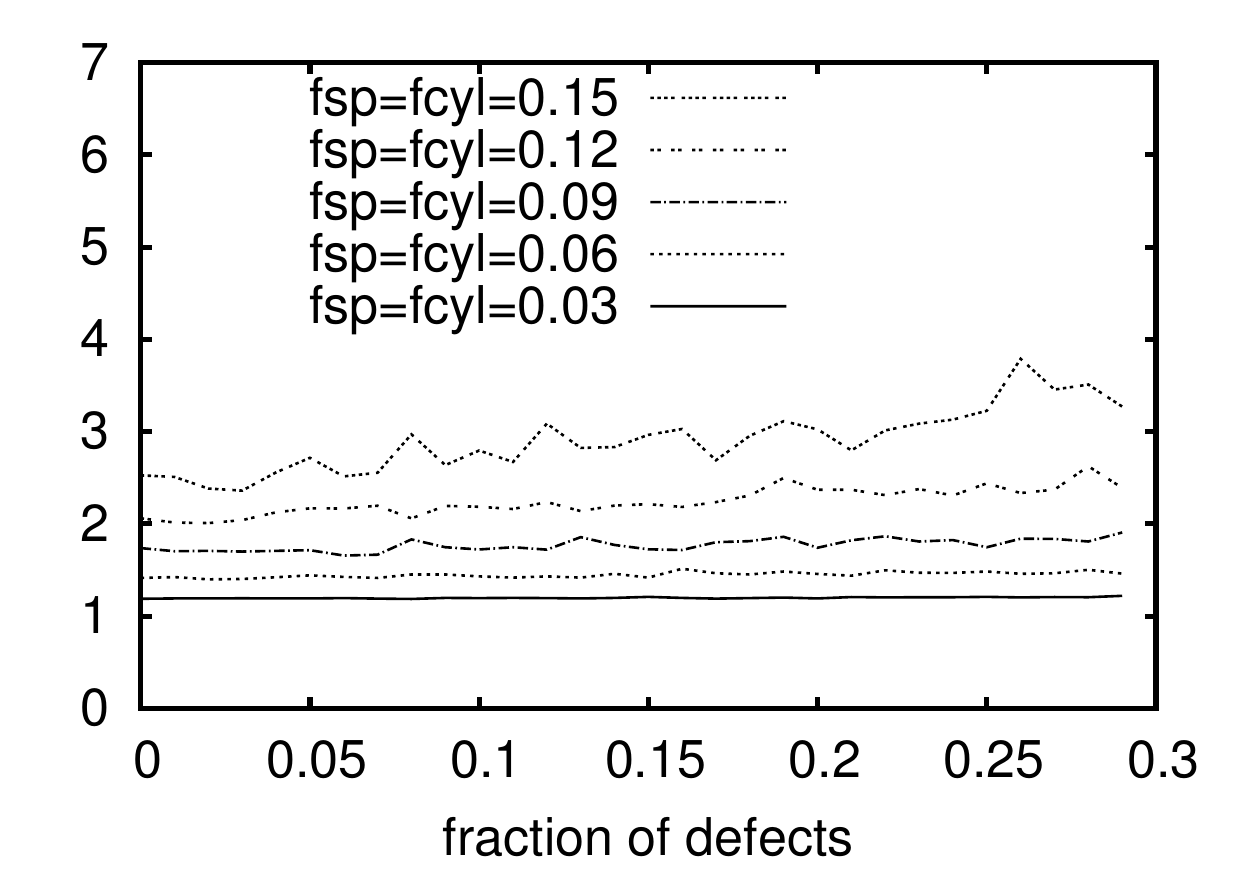} 
}
\subfigure[\, aspect 6, normalized bulk modulus]{    
 \includegraphics[width=0.45\linewidth]{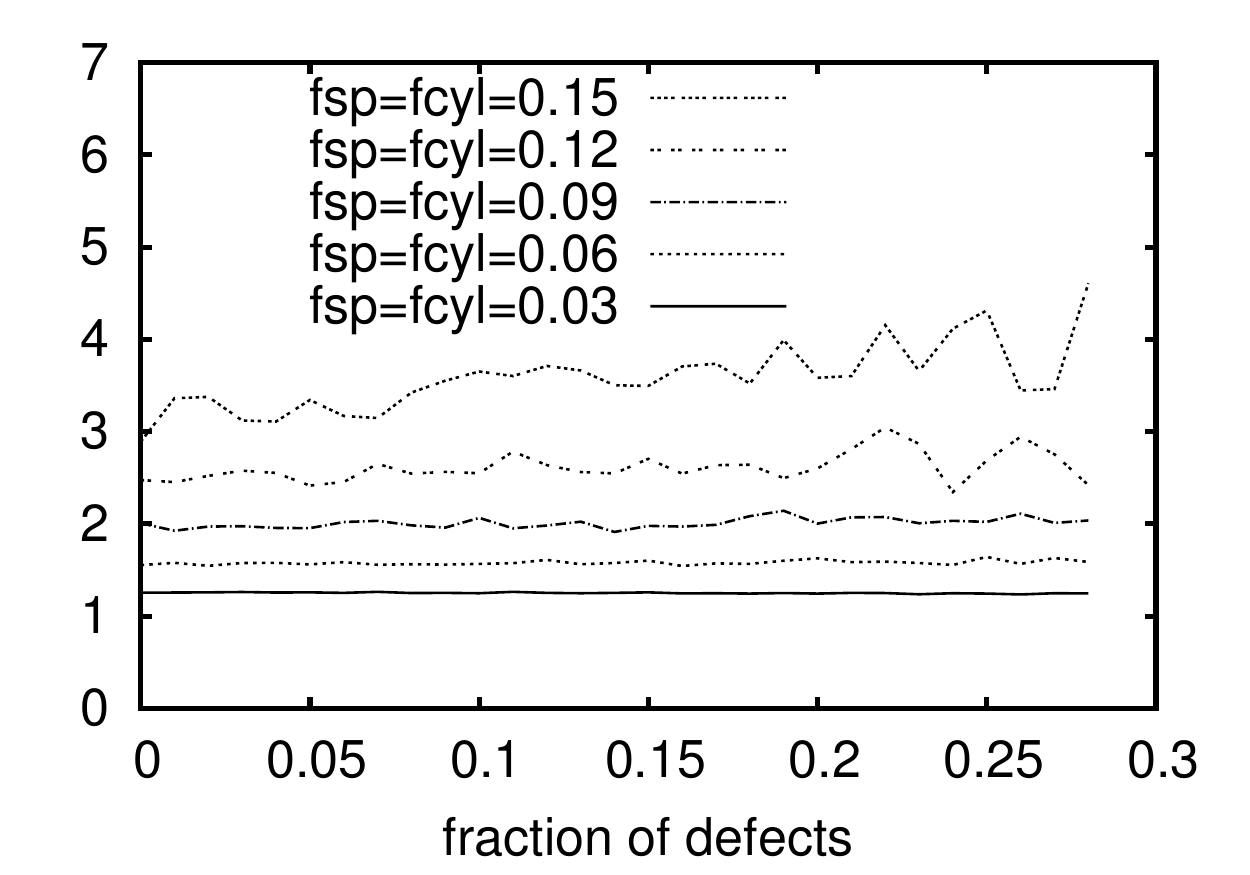} 
}

\subfigure[\, aspect 9, normalized bulk modulus]{    
 \includegraphics[width=0.45\linewidth]{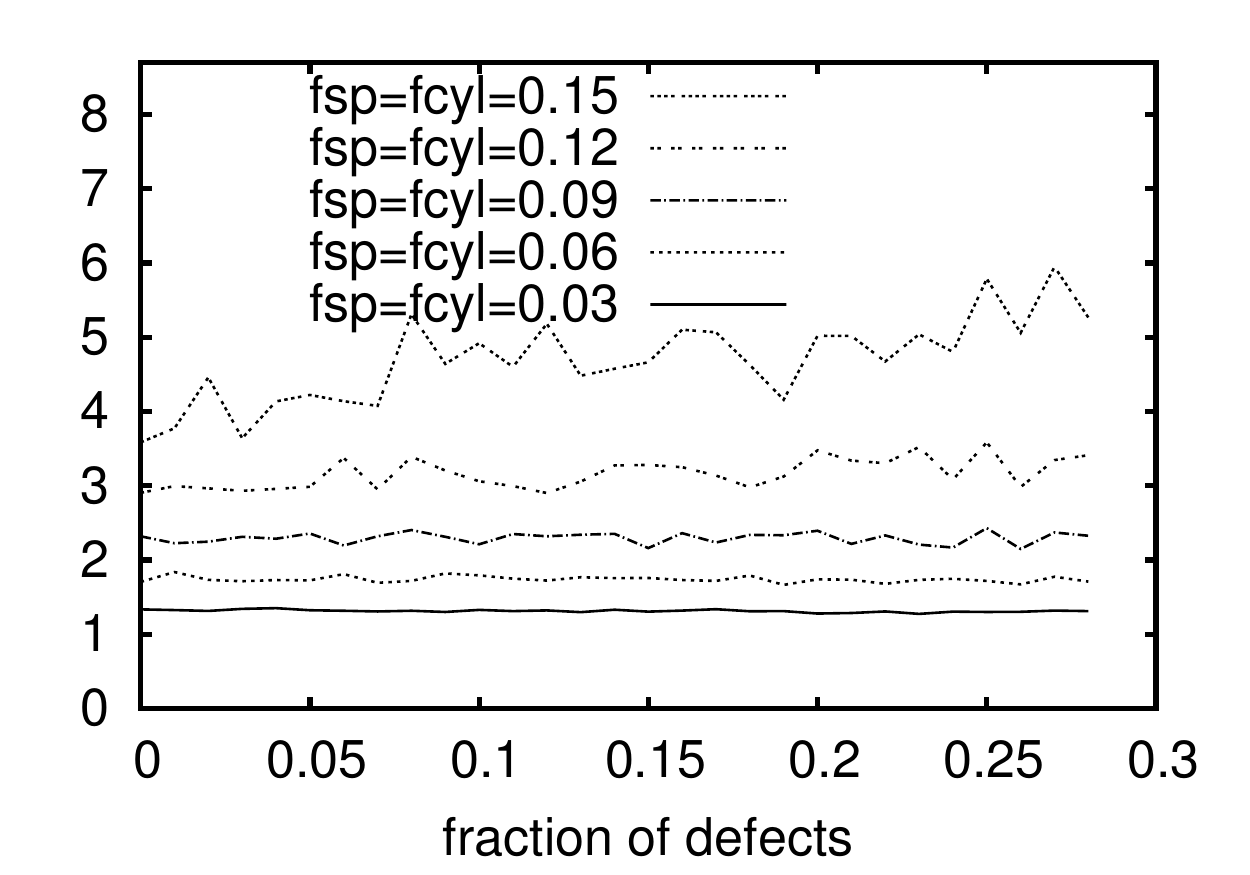} 
}
\subfigure[\, aspect 9, normalized shear modulus]{    
 \includegraphics[width=0.45\linewidth]{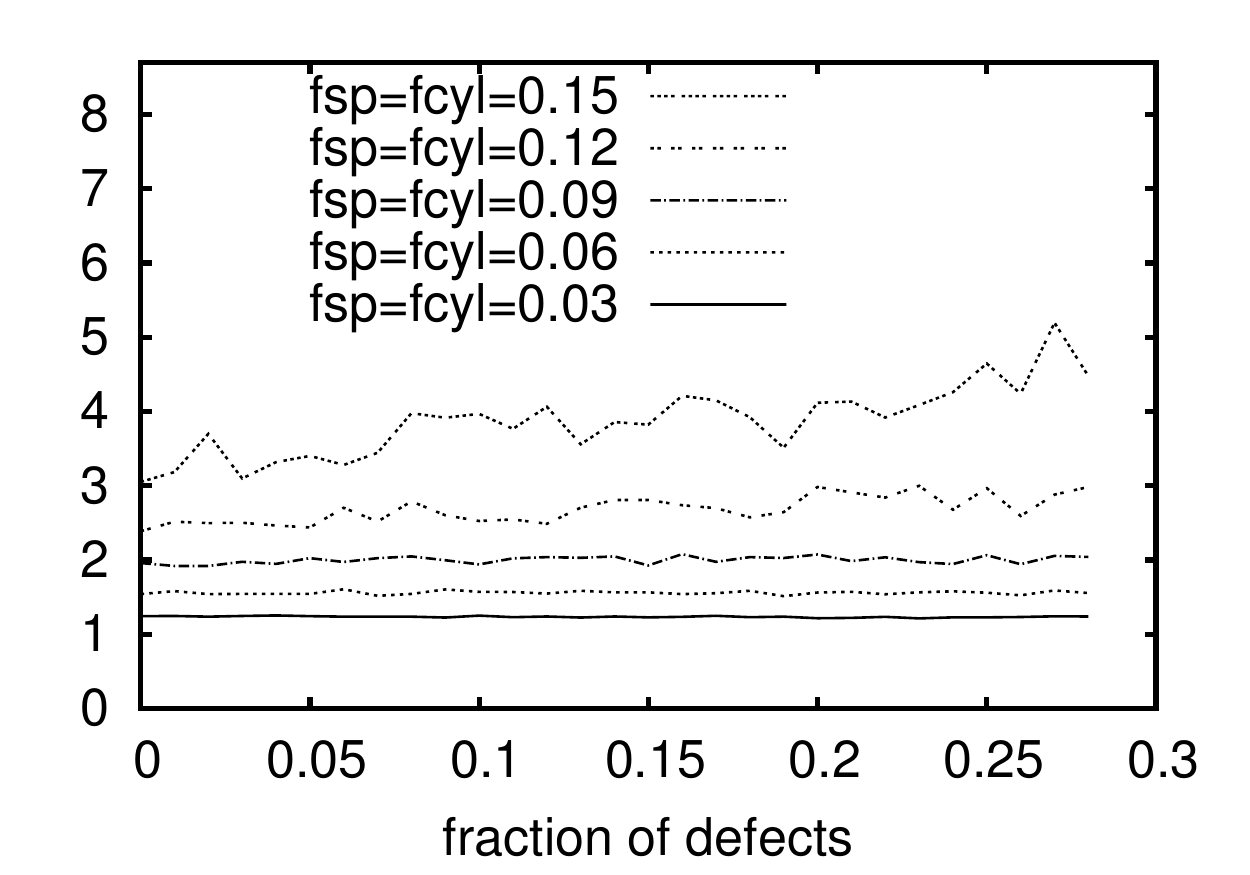} 
}

\caption{\label{fig:spoilt} The dependence of mechanical parameters of a composite
on the volume of defects in the material. $n_{sp}=n_{cyl} = 20$, contrast 2048.
} 
 \end{figure}

The second series of tests concerns the simulation of possible defects while 
introducing the inclusions to the matrix of the future composite material.
We model this process by generating the zones where the obtained composite is 
``spoilt'', namely if the inclusion intersects partially with such a zone 
some piece of it is moved out and placed apart in the matrix. The main parameter of this perturbation 
is thus the volume fraction of such zones. The figure \ref{fig:spoilt} shows the 
effect of the amount of defects on the properties of a material. 
The curves are less smooth than for most of the dependencies presented above, 
but globally one sees that the defects (at least at reasonable volume fraction)
contribute to reinforcement as well. And as before, the more the studied material was reinforced 
without imperfections, the clearer this effect is visible. 
  
% Although the trends observed in the last two series of tests may seem to be 
% counterintuitive and even surprising, they have been confirmed by our colleagues 
% doing experimental measurements for composite materials based on polymers.\footnote{This information has been 
% communicated to us during the meetings of the ACCEA industrial project.} 

\section{Conclusion / outlook}
To conclude, let us recapitulate the main messages of this paper. 
We have started by presenting the methods that we find suitable 
for analysis of mechanical properties of composite materials. The approach that 
we find optimal can be called `FFT-based stochastic homogenization', 
where the FFT-based iterative scheme is used for computing effective 
properties of each given sample, and the stochastic part enters at the 
level of random generation of samples. One of important advantages of this 
approach is its low time and memory consumption in comparison for example 
with finite elements methods. 
Turning to concrete computational results, we can conclude that the most 
important morphological properties influencing the effective properties
of composite materials are the volume fractions of various types of inclusions 
and rather basic geometry of fiber reinforcements. 
Moreover introducing various irregularities to inclusions, and thus making their 
shape geometrically more complex, has positive effect on the parameters of the 
obtained material.

We have also mentioned that behind this paper there is a precise motivation 
related to industrial applications. We are doing this work within the 
framework of an industrial project related to amelioration of effective properties 
of composite materials. Certainly the dream of an applied mathematician in such 
a context would be to propose an optimal scheme of production. Our contacts with 
the companies that actually produce composite materials show that in reality 
the fabrication process is not that flexible as one wants. 
For example, in this paper we deliberately omitted the analysis of the influence of collective 
orientation of the inclusions and their possible unequal distribution 
in a sample (though the algorithms from \cite{VDP} permit us to control 
these parameters), on the contrary we always checked that the resulting 
homogenized medium was sufficiently isotropic. 
Often such modelling results are aimed at validation of the properties 
rather than on the production planning, or at the choice 
between a very limited number of possible strategies. 
In our particular case the input  of the 
simulation consists of the parameters of different phases of the composite and
the geometry of the sample in the form of 3D images obtained 
from tomography or microscopy; and the expected outcome is the estimation 
of the effective parameters.  In the best scenario these images are segmented, 
i.e. the regions belonging to the matrix or to inclusions are labeled -- in such 
a format the sample is perfectly suitable for performing the FFT-based computation
described above. 
This however does not mean that the main applied aim of this paper was to test the methods. 
In fact during the computations described above we have accumulated a huge database 
of samples together with already computed effective properties -- this can now be 
used for various parameter fitting or inverse problems.

\textbf{Acknowledgements.} Most of the computations described in this paper have been carried out 
at the cluster of the Center of Informatics Resources of Higher Normandy
(CRIHAN -- Centre de Ressources Informatiques de HAute-Normandie).

This work has been supported by the ACCEA project selected by the ``Fonds Unique Interminist\'eriel
(FUI) 15 (18/03/2013)'' program.

%\newpage

\end{document}